\documentclass[]{elsarticle}
\usepackage{hyperref}

\usepackage{epstopdf}
\usepackage{epsfig}
\usepackage{graphicx}
\usepackage{subfigure}
\usepackage{dcolumn}
\usepackage{amsmath,amssymb}

\usepackage{upgreek}

\biboptions{sort&compress}

\makeatletter
\def\ps@pprintTitle{%
   \let\@oddhead\@empty
   \let\@evenhead\@empty
   \def\@oddfoot{\reset@font\hfil\thepage\hfil}
   \let\@evenfoot\@oddfoot
}
\makeatother

\begin{document}

\sloppy

\begin{frontmatter}

\title{Studying synthesis confinement effects on the internal structure of nanogels in computer simulations}

\author[mymainaddress]{Elena~S.~Minina}
\author[mysecondaryaddress]{Pedro~A.~S{\'a}nchez\corref{mycorrespondingauthor}}
\ead{r.p.sanchez@urfu.ru}
\author[mymainaddress]{Christos~N.~Likos}
\author[mymainaddress,mysecondaryaddress]{Sofia~S.~Kantorovich}

\cortext[mycorrespondingauthor]{Corresponding author}

\address[mymainaddress]{Faculty of Physics, University of Vienna, 5 Boltzmanngasse, Vienna, 1090, Austria}
\address[mysecondaryaddress]{Ural Federal University, 51 Lenin av., Ekaterinburg, 620000, Russian Federation}

\begin{abstract}
We study the effects of droplet finite size on the structure of nanogel particles synthesized by random crosslinking of molecular polymers diluted in nanoemulsions. For this, we use a bead-spring computer model of polymer-like structures that mimics the confined random crosslinking process corresponding to irradiation- or electrochemically-induced crosslinking methods. Our results indicate that random crosslinking under strong confinement can lead to unusual nanogel internal structures, with a central region less dense than the external one, whereas under moderate confinement the resulting structure has a denser central region. We analyze the topology of the polymer networks forming nanogel particles with both types of architectures, their overall structural parameters, their response to the quality of the solvent and compare the cases of non-ionic and ionic systems.
\end{abstract}

\begin{keyword}
Nanogels, randomly crosslinked polymer networks, computer simulations
\end{keyword}

\end{frontmatter}

\section{Introduction}

Nearly $70$ years have passed since the concept of microgel appeared for the first time as a new type of polymer system\cite{Baker1949}, in which a gel---\textit{i.e.}, a diluted network of permanently crosslinked polymers---is synthesized with the volume of a colloidal particle---\textit{i.e.}, particles with typical sizes ranging from few tens of nanometers to several micrometers \cite{2011-fernandez-nieves-bk, 2017-hamzah-jpr}. These soft particles have very interesting properties derived from their internal semiflexible network structure. For instance, they can be penetrated by other smaller particles or molecules and frequently exhibit a very strong response to external stimuli, being able to shrink or swell with a change of several times their average volume. This makes them promising building blocks of smart materials, which has greatly stimulated in recent years the research on their synthesis techniques, experimental characterization and theoretical modeling. Nowadays, it is possible to create colloidal gel particles responsive to a large variety of stimuli, including temperature, pH of the solvent or external fields~\cite{Hoare2004,Gorelikov2004,Zhang2004a,Das2006,Yin2008,Bayliss2011,2011-malmsten-ch,Holmqvist2012,Mohanty2015,Kobayashi2016,Backes2017a,Backes2017b,Colla2018}. Prospective applications are as diverse as targeted drug delivery, oil spill recovery or sensing and smart coating technologies~\cite{Raquois1995,Saatweber1996,RETAMA2003,Guo2005,LOPEZ2005,2011-malmsten-ch,Sivakumaran2011,2011-thorne-cps, Hu2012,Bonham2014,Son2016,Aliberti2017,Pepe2017,Schimka2017,Agrawal2018,Alhuraishawy2018,PU2019}. Even the term `microgel' is still applied often to gel particles of any size up to roughly 100 $\upmu$m, a conventional distinction between nanogels---referred to particles with sizes from 1 to 100 nm---and microgels---naming only particles from 0.1 to 100 $\upmu$m---is becoming widespread \cite{2007-aleman-pac}. This is favored by the fact that some important applications strictly require the use of nanometric gel particles. For instance, acting as drug delivery systems, only nanogels can overcome the blood-brain barrier \cite{Vinogradov2010}.

To date, one of the most fruitful aspects of the research on colloidal gel particles is the large amount of techniques developed for their synthesis. The most common approaches are based on the polymerization of monomers in solution in presence of crosslinking agents. The polymerization and crosslinking may take place either in homogeneous solutions, when the newly formed colloidal particles are insoluble and precipitate, or inside droplets of solution emulsions, usually in presence of surfactant agents \cite{2011-fernandez-nieves-bk, 2017-hamzah-jpr}. In the case of polymerization/crosslinking in emulsions, the droplets act as a finite size confinement for the reactions, determining the final size of the colloidal gel particle. In this way, both nano- or microgel particles can be obtained from the same polymerization/crosslinking process by only selecting the size of the emulsion droplets. Analogously, inter- and intramolecular crosslinking into colloidal particles can be also performed on already formed polymer molecules diluted in homogeneous solutions or emulsions. In most cases, molecular crosslinking is performed by means of photoinduced creation of radicals or electrochemical methods \cite{2016-mavila-chrv, 2016-galia-eccm}. These latter approaches have the advantage of leaving aside the use of free monomers and crosslinking agents, whose residual presence after the synthesis of the colloidal particles might prevent their use for some applications, particularly in biomedicine \cite{2008-kwon-oh-pps}. The current development of novel synthesis approaches aims mainly at controlling the characteristic size and dispersity of the particles, their stability and the internal distribution of every type of functional groups. Important examples of such novel techniques are the ones based on microfluidic devices \cite{Zhang2007,Shah2008,Tumarkin2009} or the use of microporous membranes for a highly controlled emulsification of the precursor solutions \cite{Charcosset2004}. The scaling down of these techniques, with the use of nanofluidic devices and emulsification with nanoporous membranes, opens up the possibility of creating nanogels with a higher control on their structure and more sophisticated properties.

Despite the significant progress achieved in synthesis techniques, the size scales involved in micro- and nanogel structures still represent a challenge for the accurate characterization and modeling of their properties. In one hand, direct experimental measurements are mainly limited to overall particle properties, as for example their size, solvent content, internal distribution of contrast agents or mechanical response \cite{2018-backes-pol}, whereas the detailed internal structure of the polymer network is still extremely challenging for direct measurements. Quantitative estimations of the degree of crosslinking of the network are usually deduced from the synthesis conditions, whereas direct observations based on conventional microscopy techniques only provide indications on the internal relative distribution of crosslinkers and other functional groups \cite{2006-hoare-jpcb}. Among conventional measurement methods, neutron and X-ray small angle scattering (SANS/SAXS) are the only ones able to provide statistical information on the internal structure and overall shape of the particles, represented as form factors. Even by combining different microscopy and scattering methods, the level of characterization detail is limited \cite{2019-witte-sm}. Only very recently, it has been possible to obtain detailed topological information on the structure of polymer networks from labelling based super-resolution microscopy measurements\cite{2016-gelissen-nnl, 2018-siemes-angw, 2018-bergmann-pccp, 2018-karanastasis-mh}. However, application of such novel technique to microgel characterization is still very scarce, whereas in the case of nanogels it still remains, to our best knowledge, unexplored. On the other hand, computer simulations are the most extensively used approach for the modeling of micro- and nanogels, being a powerful theoretical tool for the search of an accurate connection between their nanoscopic structure and their overall properties. However, these systems are complex enough to make their full explicit atomistic modeling unfeasible. Therefore, computer modeling relies on a coarse-grained representation of the polymer network, usually based on bead-spring models \cite{Grest86}. Simplest bead-spring models of nano- and microgels are based on regular lattices and have been extensively used for studying swelling properties and thermosensitivity of microgels, especially of that ones based on charged polymers \cite{Kobayashi2014,Ghavami2016,Kobayashi2016,Ahualli2017,Ghavami2017,Kobayashi2017,Hofzumahaus2018}. However, since the actual topology of the experimental polymer networks in nano- and microgels is expected to be rather complex, several efforts to move away from regular lattice architectures in order to capture such complexity have been made in last years. One of such approaches is the one presented by Gnan and co-workers \cite{Gnan2017}. Their model uses the self-assembly properties of patchy colloids in order to mimic the polymerization/crosslinking process of monomers diluted in droplet emulsions \cite{Gnan2017}, generating microgel disordered structures. This model has been used to study the internal structure and swelling behavior of microgels depending on their degree of crosslinking and size of the confining droplet \cite{Rovigatti2018}, as well as the effects of explicit solvent-polymer interactions \cite{2018-camerin-scrp}. Another model, aimed at a realistic representation of the polymer network of microgels synthesized by crosslinking of polymer molecules under confinement, has been also introduced very recently by Moreno and Lo Verso \cite{Moreno2018}. In this case, their model assumes the presence of macromolecular precursors that have been functionalized for their intermolecular crosslinking at spots with prescribed distributions, obtaining a network with a low fraction of crosslinks that has faster deswelling kinetics than equivalent microgels with regular lattice structure. Further details on the progress of numerical modeling of colloidal gel particles can be found in two very recent reviews \cite{2019-rovigatti-sm-rev, 2019-martin-molina-jml-rev}.

Independently from the approaches mentioned above, recently we introduced a model for colloidal gel particles that also intends to mimic the realistic crosslinking of polymer molecules confined in droplets \cite{Minina2018}. In difference with the model of Moreno and Lo Verso, in our case no localized functionalization of the polymer precursors diluted inside the droplet is assumed and polymer monomers are crosslinked randomly according to their proximity after equilibration. This corresponds to an experimental photonically- or electrochemically-induced crosslinking process, which, to our best knowledge, has not been represented by any other computer model to date. In our first work, we analyzed the effects of the presence of magnetic nanoparticle inclusions on the structure of the polymer network. Here, we use such model to study conventional, nonmagnetic systems.

In this work, we focus on the computer simulation study of the internal structure of simple nanogel particles synthesized by random crosslinking of polymer molecules diluted in nanodroplets, analyzing how it is affected by the relative size of the confining droplet, the quality of the solvent and the use of either non-charged polymers or polyelectrolytes as precursors. To our best knowledge, this combination of aspects has not been studied to date, neither theoretically nor experimentally. For instance, the effects of the size of the confining droplets on the synthesis of gel microparticles have been addressed in several experimental works, particularly for microfluidic synthesis approaches \cite{Zhang2007, Tumarkin2009, Crassous2015, DiLorenzo2016}. In computer simulations, as pointed above, such aspect has been also addressed only for gel microparticles obtained from different synthesis routes \cite{Rovigatti2018, Moreno2018}. However, it is reasonable to expect the effects of confinement on the crosslinking process to become more important as its size decreases, having a much higher impact on the internal structure of gel nanoparticles than in microgels. Our simulations support this hypothesis, providing indications of a non-trivial dependence of the structure on the relative droplet size when it compares to the contour length of the molecular precursors. As a consequence of such effects, we found that, while keeping the typical overall characteristics of colloidal gel particles, nanogels may exhibit internal structures different from the ones predicted in previous works for microgels.

The paper is organized as follows: in next section, we introduce the simulation model and protocol; next, we discuss the simulation results, first considering the case of non-ionic systems under good solvent conditions, followed by the analysis of the effects of solvent quality and electrostatic interactions for the case of ionic systems; finally, we conclude with a summary of results and outlook.

\section{Nanogel model and simulation method}\label{sec:methods}
In order to model the structure of nanogel particles formed from a random crosslinking of polymer molecules diluted in a nanodroplet, we adopt a coarse-grained approach based on a bead-spring representation of the polymers \cite{Grest86}, simple interaction potentials and a system of reduced units. As usual in this type of approach, the polymers are modeled as linear chains of $M$ monomers, that are simple spherical beads with reduced mass $m=1$ and reduced characteristic diameter $\sigma=1$. The excluded volume interactions of such monomers are represented by a truncated and shifted Lennard-Jones pair potential, also known as Weeks-Chandler-Andersen (WCA) potential~\cite{1971-weeks}:
\begin{equation}
U_{\mathrm{{WCA}}}(r; \epsilon, \sigma, r_{\mathrm{cut}})= \left\{ \begin{array}{ll}
U(r; \epsilon, \sigma)-U(r_{\mathrm{cut}}; \epsilon, \sigma), & r<r_{\mathrm{{cut}}}\\
0, & r\geq r_{\mathrm{{cut}}}
\end{array}\right. ,
\label{eq:WCA}
\end{equation}
where $U(r; \epsilon, \sigma) = 4 \epsilon \left [ (\sigma / r )^{12} - (\sigma/r)^6\right ]$ is the conventional Lennard-Jones potential and $r$ is the center-to-center distance between the interacting beads. Under good solvent conditions, the truncation distance is set to $r_{\mathrm{cut}}=2^{1/6}\sigma$ in order to make the interaction purely repulsive. We take the energy scale of this interaction, $\epsilon$, as well as the energy of the thermal fluctuations in the system, $kT$, as unity, $\epsilon=kT=1$. When studying poor solvent conditions, the cut-off distance for potential (\ref{eq:WCA}) acting on any pair of beads is changed to $r_{\mathrm{cut}} = 2.5\sigma$ in order to make it attractive, setting a varying attraction strength $\epsilon_a$.

The bonds between adjacent monomers along the polymer chains are represented by finitely extensible nonlinear elastic (FENE) springs \cite{Grest86},
\begin{equation}\label{eq:FENE}
U_{\mathrm{FENE}}(r; \epsilon_f, r_f)=-\frac{1}{2}\epsilon_f r_f^2 \ln\left[1-\left(\frac{r}{r_f}\right)^2\right],
\end{equation}
for what we take $r_f=1.5\sigma$ as the maximal bond extension and $\epsilon_f=22.5\epsilon/\sigma^2$ as the bond strength. The purpose of these choices is simply to keep an average distance between bonded nearest neighbors close to unity under all the conditions we sampled, preventing bond crossing while not requiring a very small integration timestep.

During the crosslinking process, the confining droplet is represented as a fixed sphere of radius $R_d$ that keeps the polymers inside its volume. This is achieved by means of the potential~(Eq.~(\ref{eq:WCA})) acting also between the polymer beads and the confining sphere, being in this case $r$ the distance from its surface to the center of the bead. Crosslinkers are simply introduced as additional bonds established between pairs of beads that were not originally bonded neighbors along the polymer chains. This is done according to the protocol described below. As a trick to improve computational efficiency, we use a different type of bonding potential for crosslinking. This is a simple harmonic bond,
\begin{equation}\label{eq:harm}
U_h (r; k_h, \sigma)=\frac{1}{2}k_h(r-\sigma)^2.
\end{equation}
By taking $k_h=10\epsilon/\sigma^2$, we checked that the mean length of the harmonic bonds after relaxation is around $1.3\sigma$ with a standard deviation of $0.2\sigma$, independently of the fraction of crosslinked monomers and the size of the nanogel, being basically equivalent to the average length of the FENE bonds. Therefore, with our choice of parameters, FENE and harmonic bonds have the same structural effects after the crosslinking process. The reason for using a different potential for the crosslinkers will be explained below, when describing the crosslinking protocol.

The case of nanogels formed by polyelectrolytes is considered by assigning electric charges to the polymer beads and introducing the Debye-H\"uckel electrostatic pair potential between them \cite{1923-debye-hueckel-pz},
\begin{equation}
U_{\mathrm{DH}}(r,q_1,q_2; \epsilon_\kappa, \kappa)=\epsilon_\kappa\frac{q_1 q_2}{r} e^{-\kappa r},\ r<r_{c},
\label{eq:DH}
\end{equation}
where $r$ is the center-to-center distance between a pair of beads 1 and 2, $q_1$ and $q_2$ their respective electric charges, $\kappa$ the Debye screening wave vector, $r_c$ the cut-off distance of the interaction and the prefactor $\epsilon_\kappa$ is defined, as is usual in this type of electrostatic calculation approach, in terms of the distance at which the electrostatic potential between two elementary charges compares to the thermal fluctuations, or Bjerrum length $\lambda_B$, so that $\epsilon_\kappa = \lambda_B kT$. The Debye-H\"uckel model is a reasonable approximation for the calculation of electrostatic interactions in polyelectrolyte systems under moderate screening, \textit{i.e.}, with a low degree of ion condensation conditions, and has been used to study the swelling behavior of nanogel particles with regular lattice internal structure \cite{Kobayashi2016}.

We choose to study nanogel particles by means of equilibrium molecular dynamics simulations in the NVT ensemble with a velocity Verlet integrator. In order to avoid the costly explicit simulation of solvent molecules, we use a Langevin thermostat. With this method, the effects of the friction and the thermal fluctuations produced by the solvent are represented implicitly by friction and stochastic terms introduced in the translational and rotational Newtonian equations of motion \cite{1987-allen, 2002-frenkel}. Since here we are only interested in equilibrium properties, we are free to choose any value for the friction constants, that we set to unity in reduced units, whereas the stochastic terms are set to satisfy the usual fluctuation-dissipation rules.

Our simulation protocol is the following. First, we fix the confining sphere representing the nanodroplet in a simulation box with open boundaries. We place inside the sphere $N_p$ equivalent polymer chains of $M$ beads each. The radius of the confining sphere is taken according to the characteristic size of an unconstrained polymer chain of such length,
\begin{equation}
 R_d=0.5\sigma M^\nu,
\end{equation}
where $\nu=0.59$ is the Flory exponent. We perform an initial relaxation of the polymer chains inside the sphere, letting them to interact only through the excluded volume and confinement repulsions defined from potential Eq.~(\ref{eq:WCA}), by integrating for $10^6$ steps with a time step of $\delta t = 0.01\tau$, where $\tau$ is the time scale in our system of reduced units, $\tau=\sigma (m/\epsilon)^{1/2}$. This is followed by the crosslinking process, that essentially consists in the selection of $N_c$ pairs of beads from different polymer chains to be bonded with potential (\ref{eq:harm}), so that the final crosslinking fraction, defined as the ratio between the amount of crosslinks and the amount of beads, is $\phi_{links} = N_c (N_p M)^{-1}$. Such selection is made by randomly picking from a list of all pairs of particles belonging to different chains that are separated by a distance not larger than an arbitrary cutoff. Initially, we set such cut-off distance to $1.2\sigma$ but, whenever it is necessary, we slowly increase it until the desired amount of crosslinks is reached. At this point, the use of a harmonic potential instead of a FENE potential for the crosslinking becomes clear: it allows to establish bonds with lengths that, initially, can be slightly larger than the maximum extension of the FENE bonds but, after equilibration, tend to relax to the same equilibrium length. This trick largely simplifies the mimicking of the random crosslinking process. To limit the crosslinking to interchain bonds is a simple way to ensure the connectivity of the structure and is a reasonable assumption at least for moderately dense dilutions of polymers, which keep a high degree of mixing under confinement \cite{Jun2007}, and/or for radiation-induced polymer crosslinking processes at room temperature~\cite{2011-an-pol}. The case when good mixing conditions are not satisfied will be discussed below. Once the crosslinks are established, the system is let to relax for $10^6$ integration steps. Finally, the confining sphere is removed and the crosslinked object, with the full set of interactions, is equilibrated and its properties are measured in a run of $10^8$ integration steps. The results presented here were obtained by averaging over 50 independent runs, \textit{i.e.}, over 50 different network configurations obtained from the crosslinking process, subsequently equilibrated for each set of parameters. The model and simulation protocol were implemented using the {ESPResSo} 3.2 simulation package \cite{arnold13a}.

\section{Results and discussion}
We sample a set of parameters specifically chosen to analyze whether confinement effects can be important in the structure and properties of nanogels made by random intermolecular crosslinking of polymer precursors. For the crosslinking process, we consider two different solutions consisting of $N_p=6$ monodisperse polymer chains with either $M=100$ or $M=200$ beads each, so that the resulting crosslinked particle will consist of either $N=600$ or $N=1200$ beads in total. According to the criterium described above, the radius of the sphere representing the confining nanodroplet is therefore taken as $R_d=7.57\sigma$ and $R_d=11.39\sigma$, respectively, corresponding to volume fractions of polymer beads of $\phi_{beads}\approx0.17$ for the smaller system and $\phi_{beads}\approx0.10$ for the larger one. Since the key point we want to address here is the effect of the degree of confinement during crosslinking and not of the size of the system, in the following we will refer to the case $N=600,\,\phi_{beads}\approx 0.17$ as the system crosslinked under relatively strong confinement conditions (SCC), and to the case $N=1200,\,\phi_{beads}\approx 0.10$ as that one crosslinked under moderate confinement conditions (MCC). Regarding the equivalence to physical quantities in both cases, assuming for example that beads in our model represent Kuhn lengths of the polymer chains, these values could correspond to poly-vinylpyrrolidone (PVP) molecules with contour lengths of 300 and 600~nm \cite{2010-knappe-pol} diluted in droplets with diameters of approximately 45 and 68~nm, respectively. Such droplet sizes are comparable to that ones of nanogel particles synthesized by pulse electron irradiation- or electrochemically-induced crosslinking of PVP molecules \cite{2011-an-pol, 2016-galia-eccm}. However, we should underline that here we aim at addressing a fundamental question rather than at the accurate modeling of a particular system, thus the chosen values could also be scaled to represent other systems of polymers with different Kuhn lengths and droplet sizes. Therefore, only ratios of the sampled values are important in the following discussion.

Experimental random crosslinking techniques make more difficult the estimation of the degree of crosslinking present in the resulting nanogel particles than in the case of conventional chemical crosslinking methods. Typical values of the fraction of crosslinks with respect to the total amount of monomers provided by the latter techniques in micro- and nanogels are around $\phi_{links}\sim0.1$, but values in a range as broad as $0.01$ to $0.8$ have been reported \cite{Bonham2014}. Since we aim at a qualitative discussion, we chose to sample three different values from such range, from very low to strong: $\phi_{links}=0.03$, $0.17$ and $0.33$. We assume all these values to be reachable by radiation- and electrochemically-induced crosslinking methods.

In the next sections, we first discuss the case of non-ionic systems under good solvent conditions, analyzing the topology of the polymer networks and their structural parameters.

\subsection{Topology}
\begin{figure}
    \centering
    \subfigure[]{\includegraphics[width=7cm]{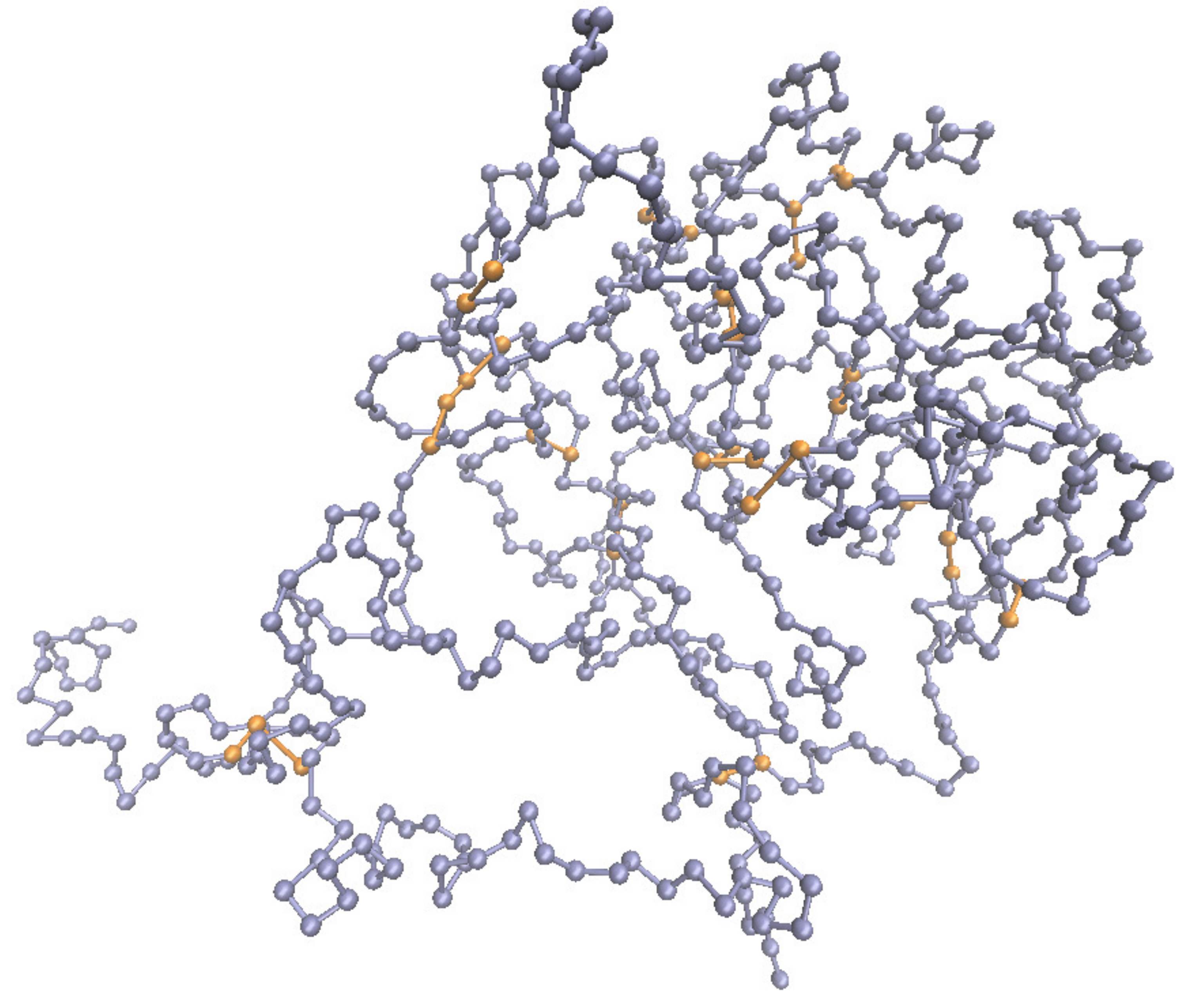}}
    \subfigure[]{\includegraphics[width=4.cm]{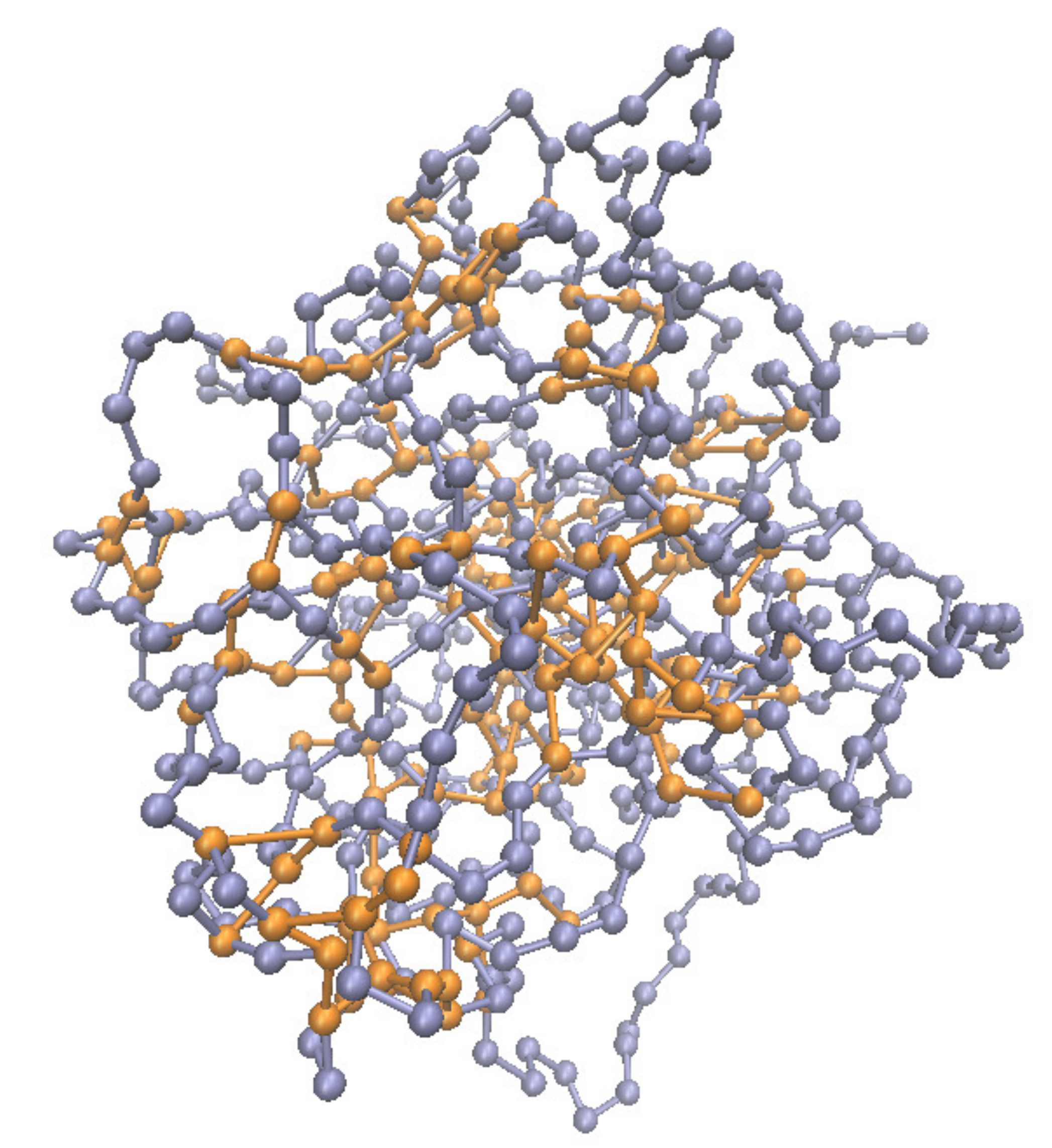}}
    \subfigure[]{\includegraphics[width=3.cm]{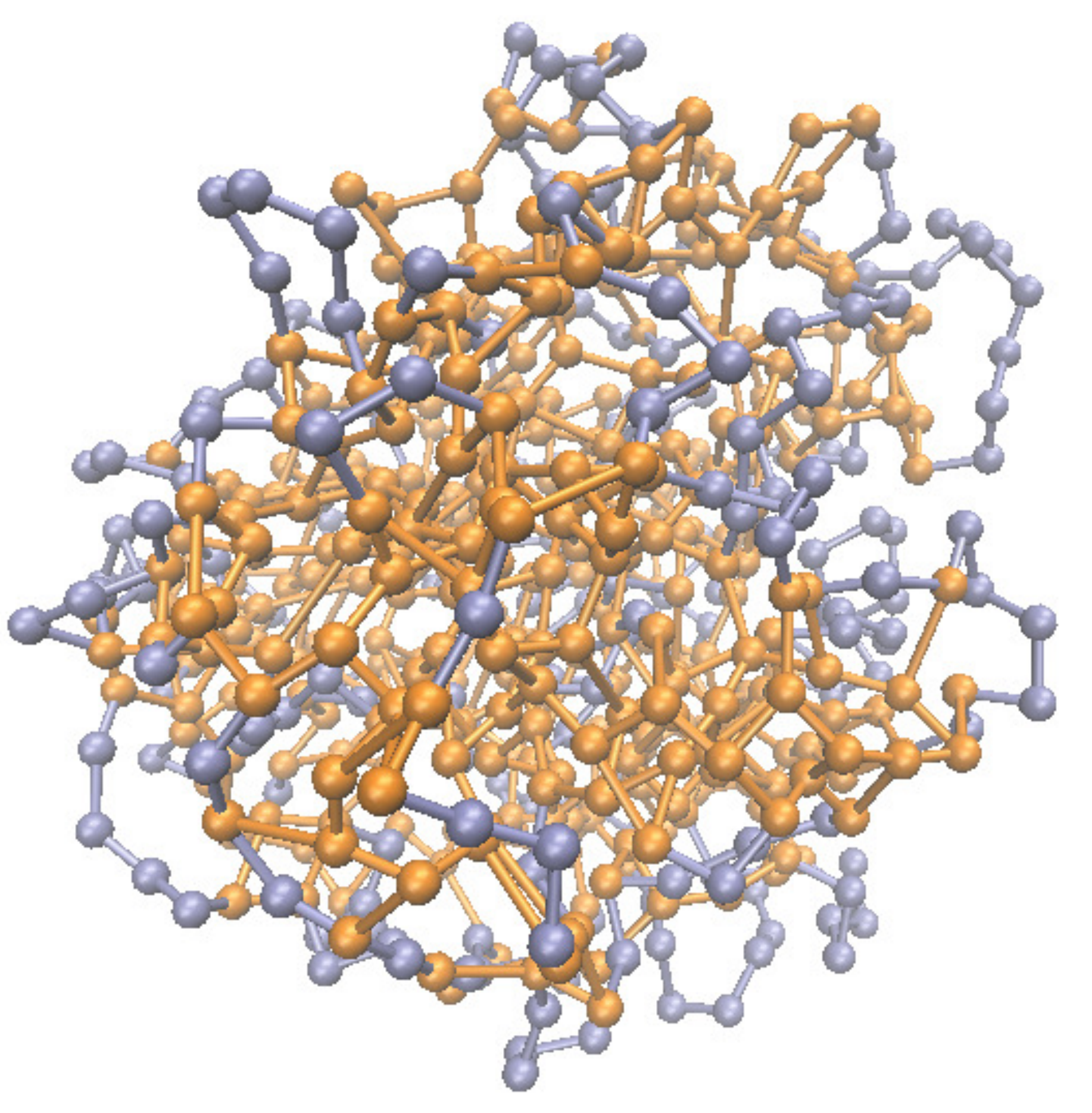}}
    \caption{Equilibrium configuration snapshots of nanogel particles created under strong confinement conditions (SCC) in a good solvent. Beads are represented as spheres and bonds as linear segments. Diameters of the beads are decreased to $0.5\sigma$ to ease the visualization. The fraction of crosslinked monomers is (a) $\phi_{links}=0.03$, (b) $\phi_{links}=0.17$, (c) $\phi_{links}=0.33$. Crosslinked and non-crosslinked beads are colored in orange and blue, respectively.}
    \label{fig:snapshots_good}
\end{figure}
We start our discussion by characterizing the structural properties of the polymer networks obtained from our random interchain crosslinking procedure. Fig.~\ref{fig:snapshots_good} shows some examples of typical equilibrium configurations of the networks obtained under good solvent conditions, corresponding to the non-ionic case and each of the sampled crosslinking fractions. As expected, one can see that the overall structure becomes more compact and spherical as the fraction of crosslinks increases. For the lowest crosslink fraction, the structure is so loose that it can hardly be considered an actual soft particle, whereas for $\phi_{links}=0.17$ and $\phi_{links}=0.33$ the configurations look similar to the ones obtained from other computer models. Therefore, with this set of crosslinks fractions we can compare the structure of a loose, small polymer network with the ones of nanogel particles.

\begin{figure}[h]
\centering
 \subfigure[]{\label{fig:P_lin_seg_600}\includegraphics[width=0.48\textwidth]{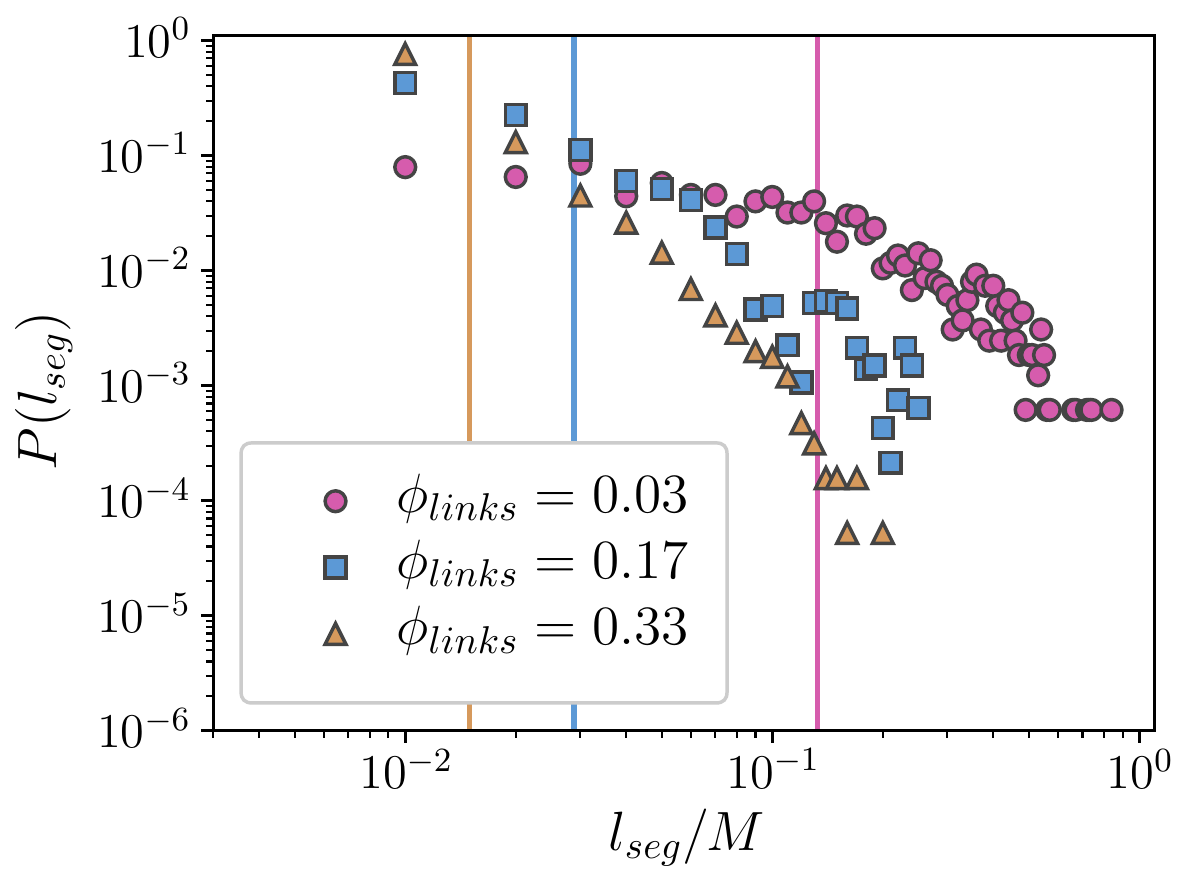}}
 \subfigure[]{\label{fig:P_lin_seg_1200}\includegraphics[width=0.48\textwidth]{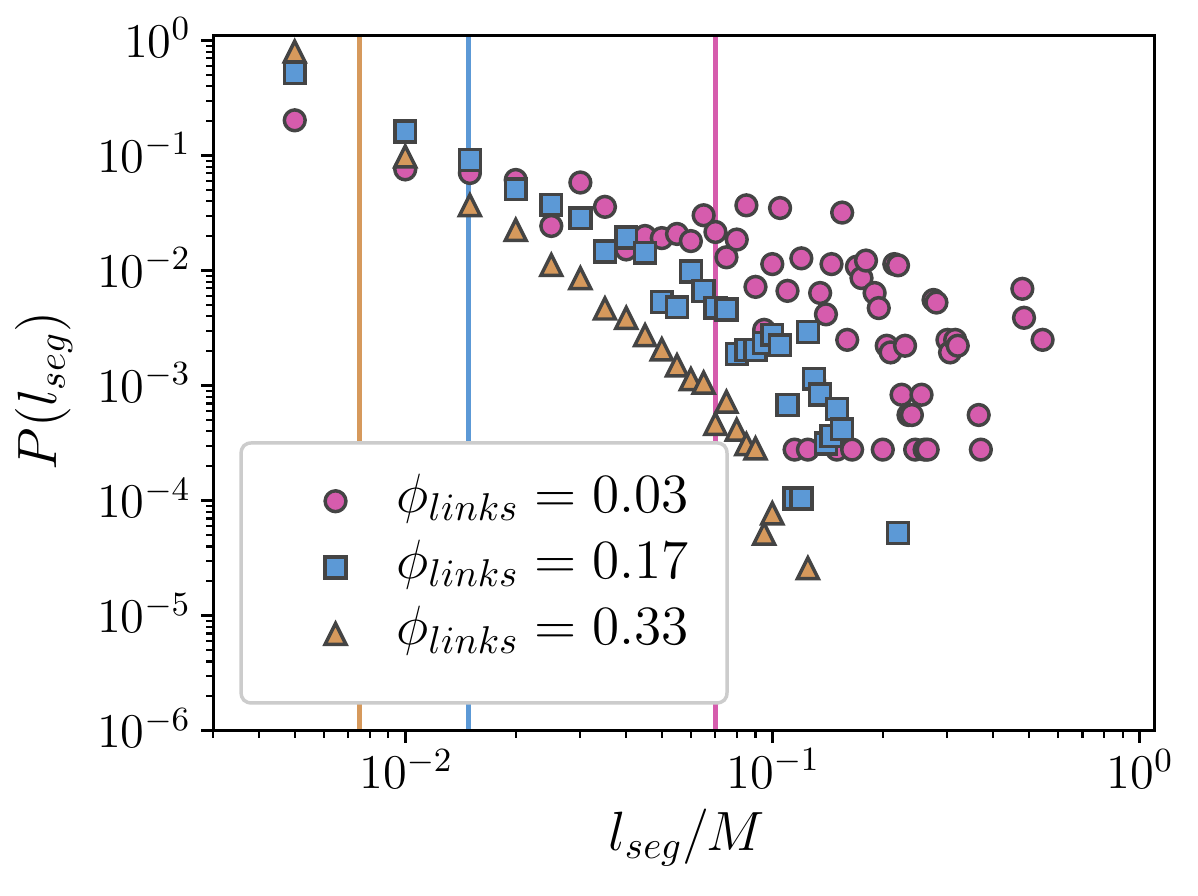}}
 \caption{Probability distributions for the length of linear segments, $l_{seg}$, in non-ionic networks equilibrated in good solvent corresponding to (a) strongly confined crosslinking (SCC) and (b) moderately confined crosslinking (MCC), at different fractions of crosslinks, $\phi_{links}$. Horizontal axis is rescaled by the corresponding individual polymer length, $M$. Vertical lines mark the relative average segment length, $\langle l_{seg}^{max}\rangle/M$. Values for increasing $\phi_{links}$ are, respectively: (a) $\langle l_{seg}^{max}\rangle /M \approx 0.133, 0.029$ and $0.015$; (b) $\langle l_{seg}^{max} \rangle /M \approx 0.070, 0.015$ and $0.008$.}\label{fig:P_lin_seg}
\end{figure}

Unlike in other simulation models, and despite the apparent resemblance in the resulting overall structure, our polymer networks have no predefined linear segment distribution. Therefore, it is interesting to analyze such parameter in our networks. A linear segment with length $l_{seg}$ is defined as a part of the polymer network consisting of a sequence of bonded but not crosslinked beads that are bounded by crosslinked beads and/or chain free ends. Since we have different bonds for the polymer backbones and the crosslinks, we can compute $l_{seg}$ easily by simply counting the number of FENE bonds in each bounded segment. The relative probability distributions of $l_{seg}$ are presented in Fig.~\ref{fig:P_lin_seg_600} for networks obtained under strongly confined crosslinking and in Fig.~\ref{fig:P_lin_seg_1200} for that ones obtained under moderate confinement.
As one can expect, for each system type, the distributions become more steep and narrow as the fraction of crosslinks increases, leading to a lower average segment length. By comparing both system types, we can see that the relative distributions of $l_{seg}$ are broader for SCC systems, showing larger relative average values, $\langle l_{seg}^{max} \rangle /M$. Regarding the precursor lengths used in each case, by doubling the length of the polymer backbones we obtain only an increase of the average segment length of approximately $6\%$ for the least crosslinked system, $3\%$ for the intermediately crosslinked system and less than $0.3\%$ for the most crosslinked one. This indicates a weak dependence of $\langle l_{seg} \rangle$ on the length of the polymer precursors, that tends to vanish as the fraction of crosslinks increases.

\begin{figure}
 \centering
 \subfigure[]{\label{fig:dens_prof_in}\includegraphics[width=0.45\textwidth]{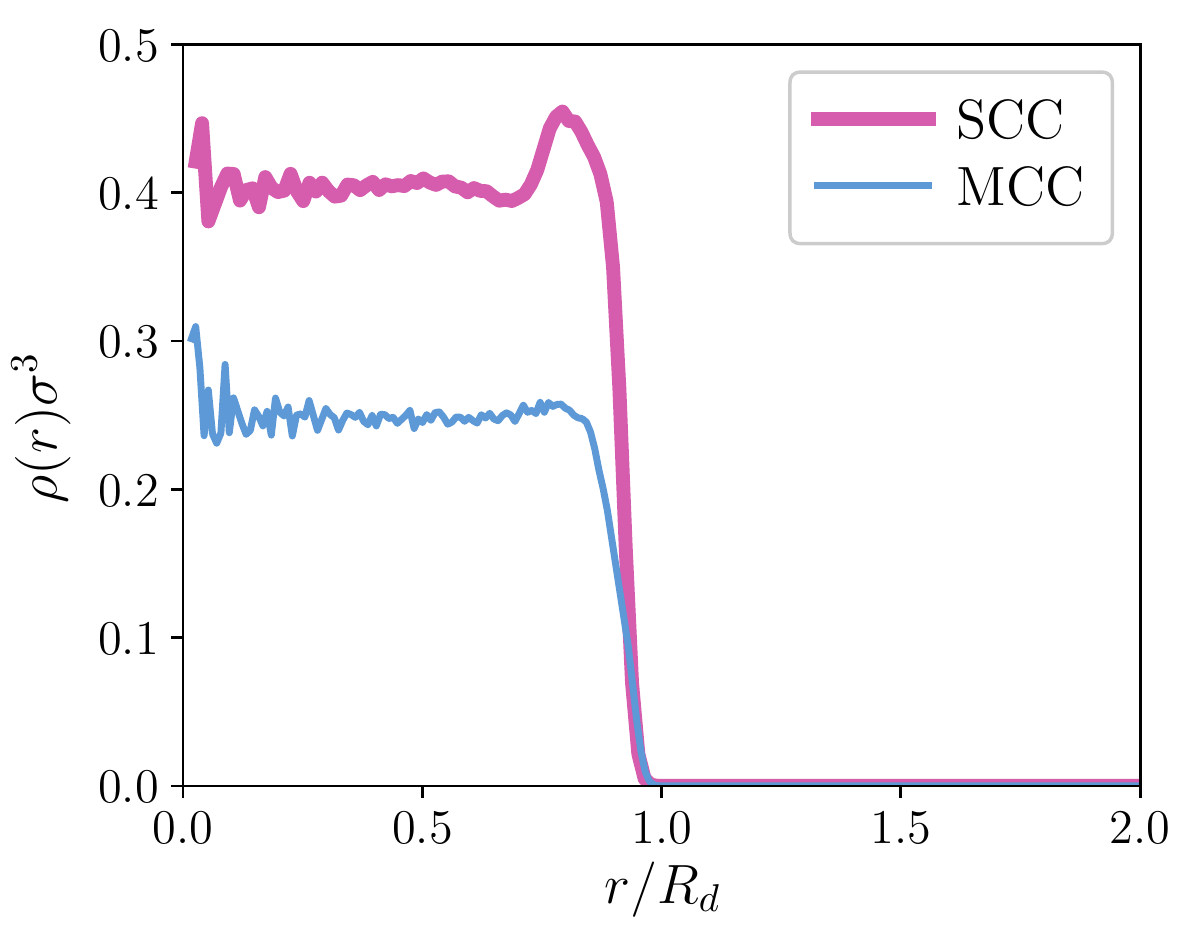}}
 \subfigure[]{\label{fig:P_cross-link_100}\includegraphics[width=0.45\textwidth]{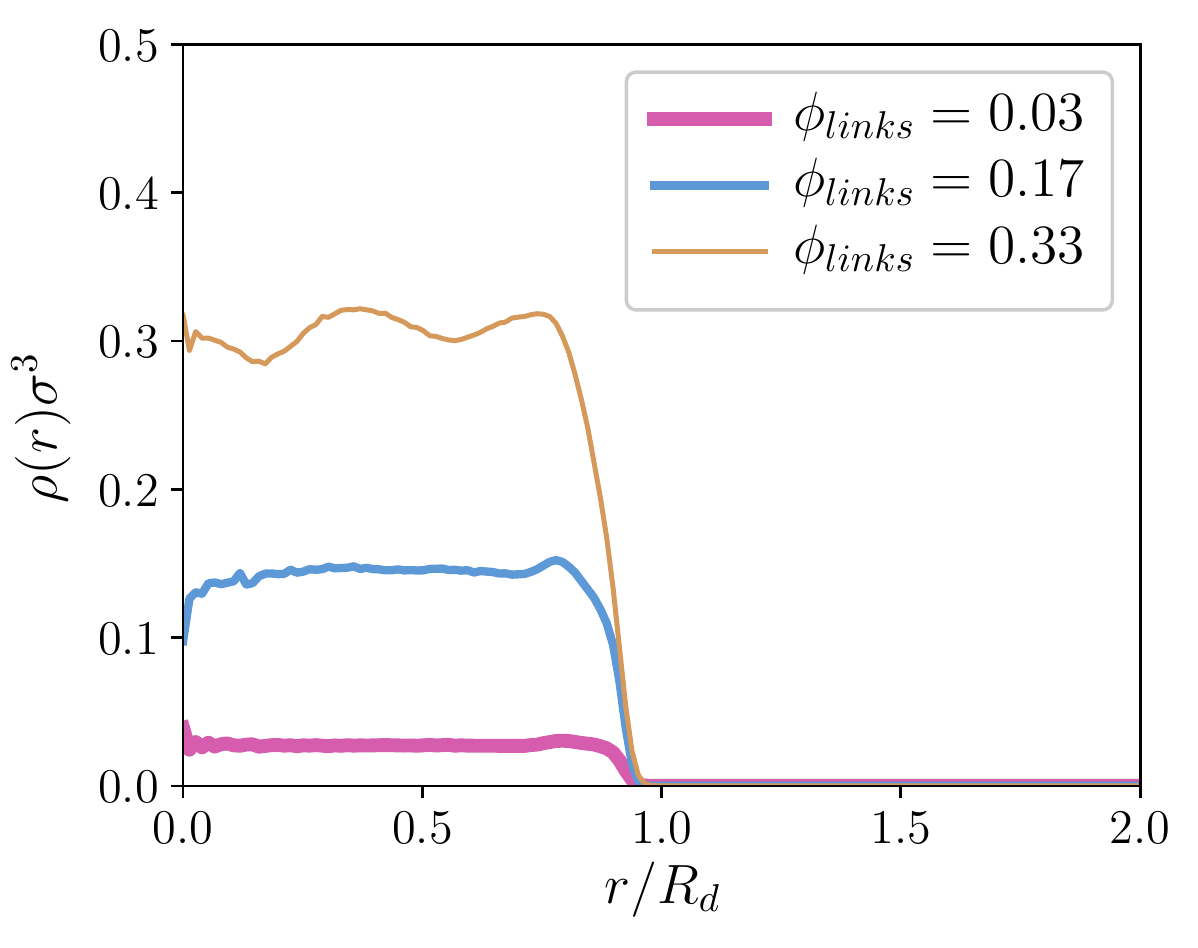}}
 \subfigure[]{\label{fig:P_cross-link_200}\includegraphics[width=0.45\textwidth]{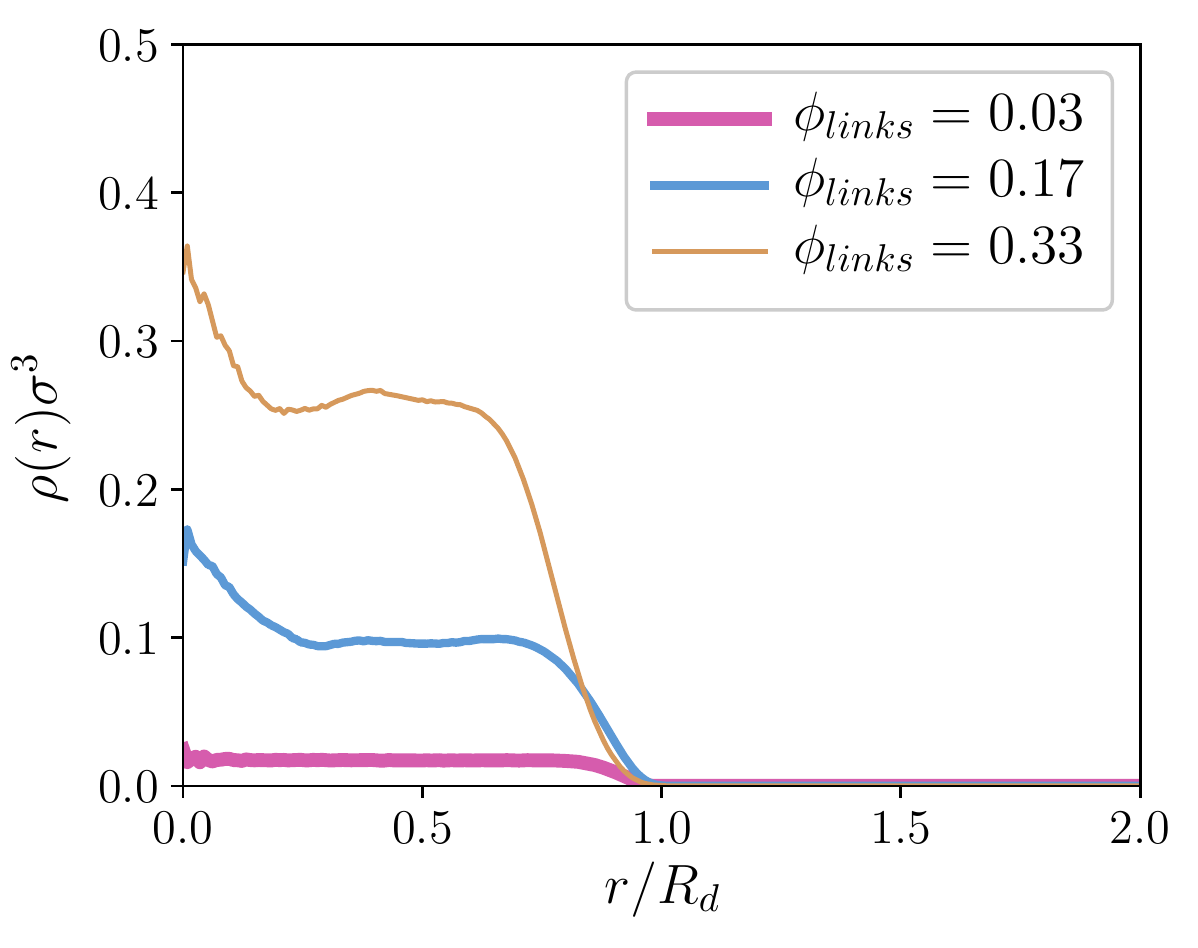}}
 \caption{ Radial number density profiles of systems remaining under spherical confinement, measured from the center of the confining sphere. (a) Density profiles of polymer beads before the crosslinking process for both sampled confinement conditions. (b) and (c): density profiles of crosslinked monomers only, obtained right after the crosslinking procedure for each fraction of crosslinks: (b) SCC system; (c) MCC system.}\label{fig:density_prof}
\end{figure}
The reason for the weak dependence of $l_{seg}$ on $M$ is not obvious and one may wonder whether it is the result of a confinement effect. In order to analyze this, in Fig.~\ref{fig:density_prof}, we present several reduced number density profiles of different equilibrated systems still under confinement, measured from the center of the confining sphere. Fig.~\ref{fig:dens_prof_in} shows the density profiles of polymer beads obtained for both sampled confinement conditions right before the crosslinking takes place. A qualitative difference can be observed: while the MCC system shows a rather uniform distribution, in the more strongly confined a maximum of density is found close to the boundary of the confining sphere, indicating that the polymers tend to occupy the external region of the droplet rather than the center. This is the signature of the system being in a highly concentrated regime, as defined for polymers under spherical confinement \cite{Jun2007}. Under such conditions individual chains can segregate at no penalty in their free energy and concentrate close to the sphere boundaries. Having a significantly lower density, the moderately confined system does not display such behavior. This difference cannot but affect the distribution of crosslinks. This is evidenced by Fig.~\ref{fig:P_cross-link_100} and \ref{fig:P_cross-link_200}, that shows the number density profiles of crosslinks for each system size and fraction of crosslinks. Without regard of the fraction of crosslinks, we can observe that in the MCC system (Fig.~\ref{fig:P_cross-link_200}) most crosslinks are located in the center of the sphere, with a relatively flat middle region that decays when approaching the boundary. However, in the SCC system, such preference for the central region is absent and a small maximum can be observed close to the boundary.

\begin{figure}[h]
 \centering
 \subfigure[]{\label{fig:snaps_100}\includegraphics[width=0.45\textwidth]{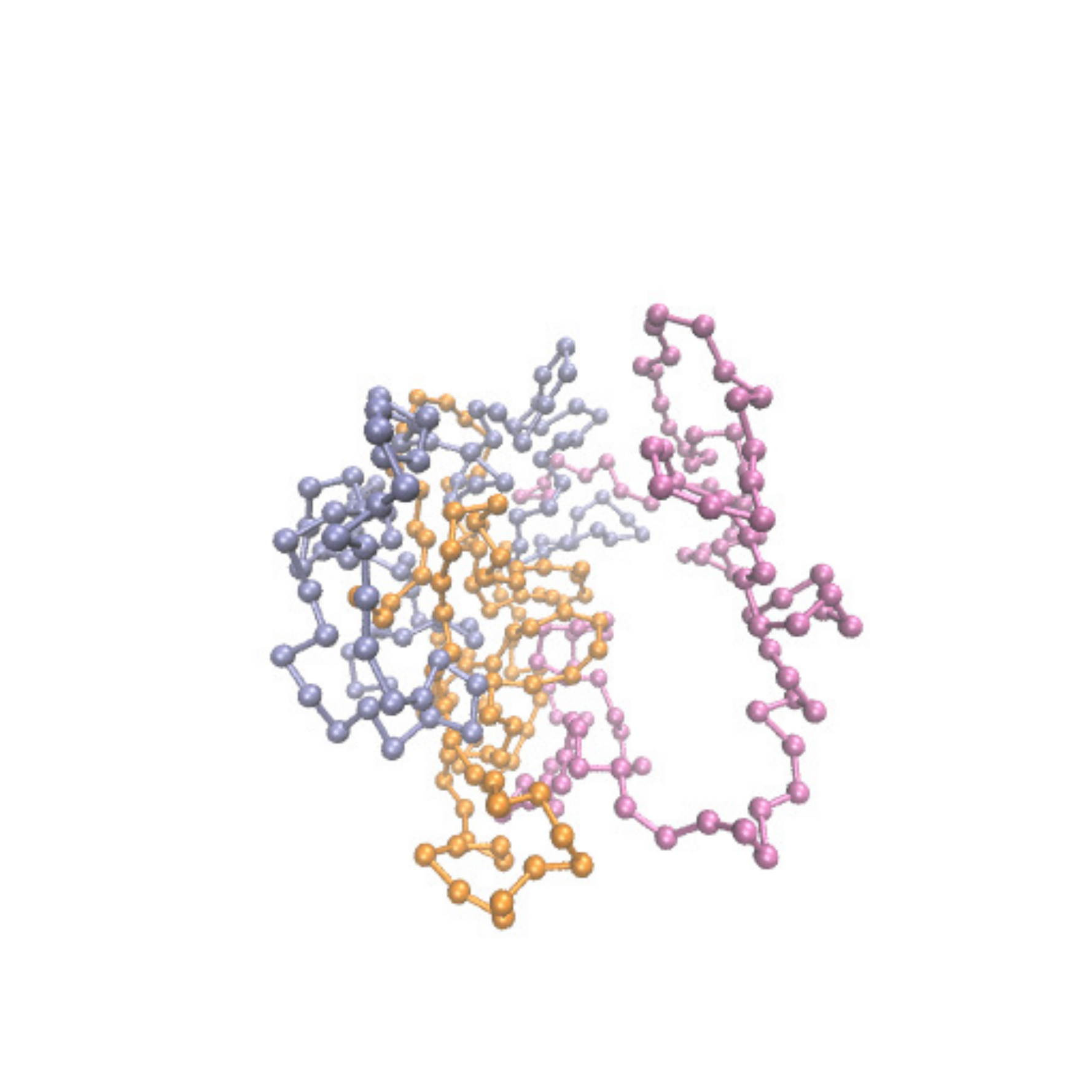}}
 \subfigure[]{\label{fig:snaps_200}\includegraphics[width=0.43\textwidth]{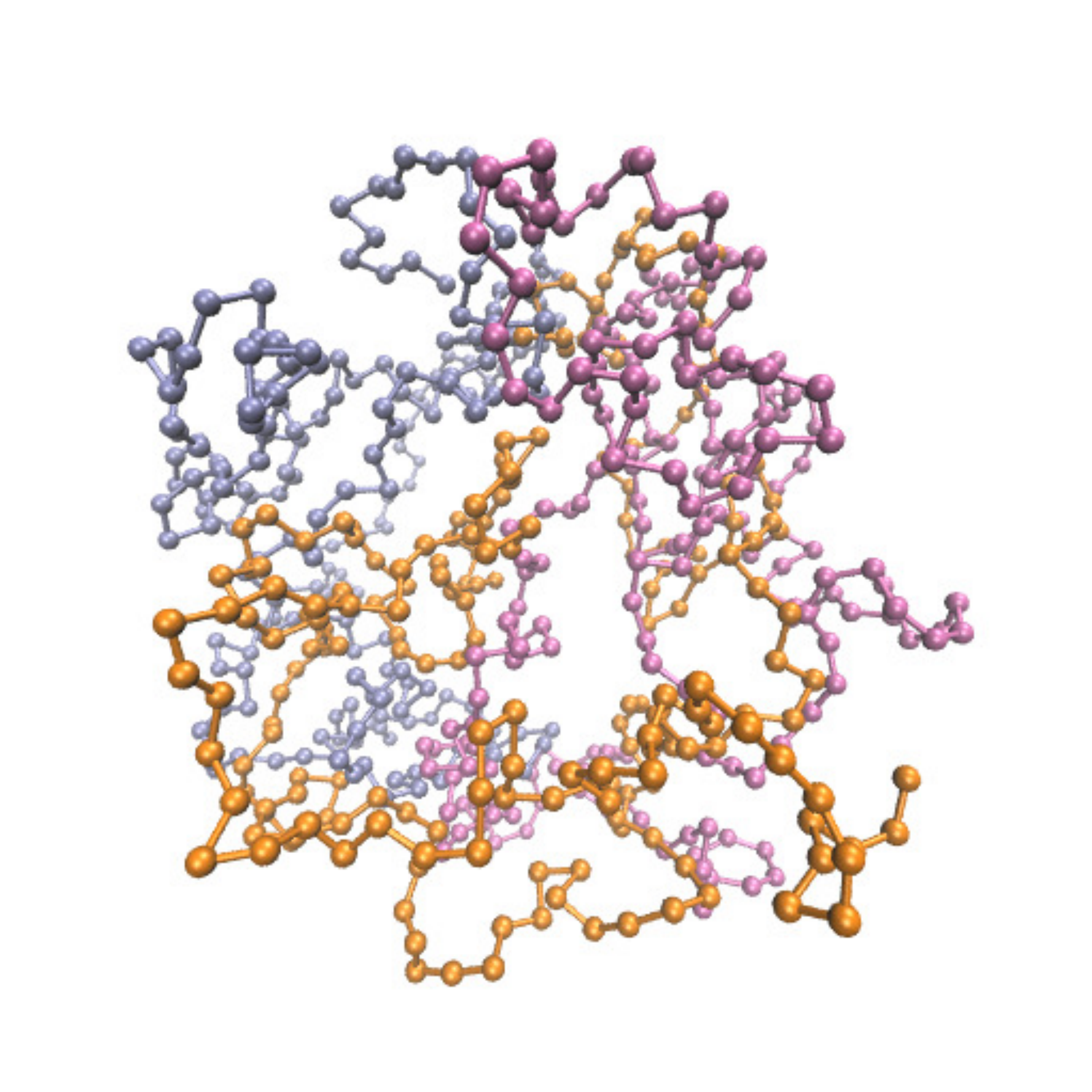}}
 \caption{Snapshots of three polymer chains belonging to systems with $\phi_{links}=0.17$ still under spherical confinement: (a) SCC system; (b) MCC system.}\label{fig:snaps}
\end{figure}
Fig.~\ref{fig:snaps} provides a further indication of the rather different structure adopted by the polymers under distinct confinement conditions. It shows two configuration snapshots corresponding to the point in our simulation protocol right before removing the spherical confinement. To ease the visualization, only three polymer chains with different colored backbones from each system are depicted. By comparing the SCC (Fig.~\ref{fig:snaps_100}) and MCC (Fig.~\ref{fig:snaps_200}) cases, we can see that the longer polymer chains of the latter system have a more uniform distribution inside the cavity than the shorter chains of the former one. In fact, shorter chains tend to be more aligned with the confinement wall.

At this point, the weak dependence of the linear segment length on $M$ can be explained by the different configurations adopted by the polymer chains when the crosslinking is performed. If chains in a solution of $N_p$ polymers of length $M$ are perfectly mixed, from simple probability considerations one can estimate that the average linear segment length after randomly assigning $\phi_{links} N_p M$ interchain crosslinks is approximately
\begin{equation}
\langle l_{seg} \rangle \approx \frac{1}{2 \phi_{links}}.
\label{eq:seglen_mixed}
\end{equation}
The values provided by this expression for the MCC system are, for increasing fraction of crosslinks, $\langle l_{seg} \rangle \approx 3,\,7$ and $17$, which are very close to the ones measured from our configurations (see values in caption of Fig.~\ref{fig:P_lin_seg}). However, for the SCC system, expression~(\ref{eq:seglen_mixed}) significantly underestimates the values of $\langle l_{seg} \rangle$ actually obtained, indicating that in this case the polymer chains were not completely mixed during crosslinking. Chain segregation in a dense system, even when it is only partial, makes individual chains to form blob-like structures. Such structures necessarily decrease the amount of nearly close contacts between different chains with respect to a system of perfectly mixed chains. Therefore, the random interchain crosslinking procedure based on a distance criterium will involve only chain segments at the interfaces of the blobs, leaving that parts in the inner region of the blobs without crosslinks. Compared to a well mixed case, this makes the distribution of $l_{seg}$ more disperse, with more long non-crosslinked segments, increasing its average value. In summary, we found two manifestations of a significant effect of strong confinement conditions on the internal structure of randomly crosslinked polymer networks and nanogel particles: the increase of the average linear segment length with respect to moderate confinement conditions and, more importantly, the adoption of a structure slightly more dense and crosslinked in the external region of the droplet than in the center.

The confinement effect discussed above is interesting and might be relevant for nanogel particles synthesized under analogous conditions. However, one may argue that this effect is only an artifact of our approximations, since here we are considering exclusively interchain crosslinking, whereas intrachain crosslinking should be significant at least for the more tightly confined system, in which polymer mixing is poor. However, intrachain crosslinking can not prevent this effect to manifest, at least in the peculiar profile of the strongly confined structure. The reasoning behind this affirmation is the following. First, in any case some degree of interchain crosslinking is necessary in order to obtain nanogel particles with larger molecular weight than each individual precursor polymer, being single molecule particles out of the scope of this work. Second, intrachain crosslinking will be still selected only by the close distance criterium, not being affected by the formation of segregated chain blobs. Therefore, the distribution of polymer beads and crosslinks will be essentially the same, with the only difference that the latter will be a mixture of both kinds of crosslinks, inter- and intrachain ones.

\begin{figure}
 \centering
 \includegraphics[width=7cm]{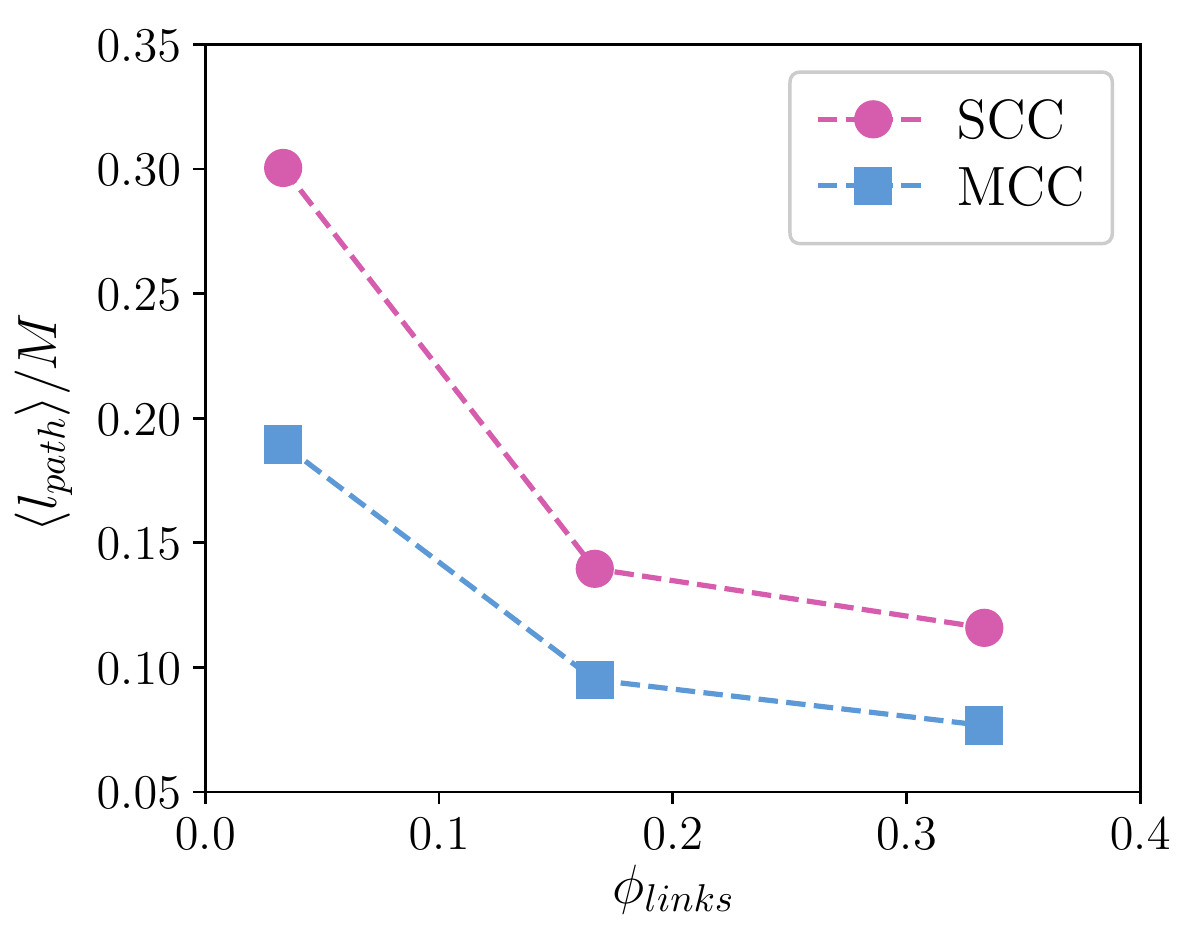}
 \caption{Average shortest path length between crosslinked beads, $\langle l_{path} \rangle$, rescaled by the length of polymer chains, $M$, as a function of the fraction of crosslinked monomers, $\phi_{links}$, for both sampled crosslinking confinement conditions.}\label{fig:short_path}
\end{figure}
Another topological parameter of interest that may evidence the effect of confinement in the structure of the crosslinked networks is the average shortest path between crosslinks, $\langle l_{path} \rangle$, defined as
\begin{equation}
\langle l_{path} \rangle =\frac{1}{N(N-1)}\sum\limits_{i,j}l_{sh}(i,j),
\end{equation}
where $l_{sh}(i,j)$ is the shortest linear segment that connects the crosslinked beads $i$ and $j$. Fig.~\ref{fig:short_path} shows $\langle l_{path} \rangle$, normalized by the polymer backbone length, as a function of the fraction of crosslinks. We can see that the average shortest path decays with $\phi_{links}$, as it is expected. Interestingly, $\langle l_{path} \rangle/M$ is approximately  $1.5$ times smaller for the MCC case than for the SCC one. This result also supports the existence of a more homogeneously crosslinked network in the less confined system than in the more strongly confined one.

\subsection{Gyration radius and asphericity}
Once we examined the differences in the topology of the polymer networks led by the confinement conditions during crosslinking, we analyze larger scale structural properties of these systems once the confinement is removed.

\begin{figure}[h]
 \centering
 \subfigure[]{\label{fig:Good_size_radius}\includegraphics[width=0.45\textwidth]{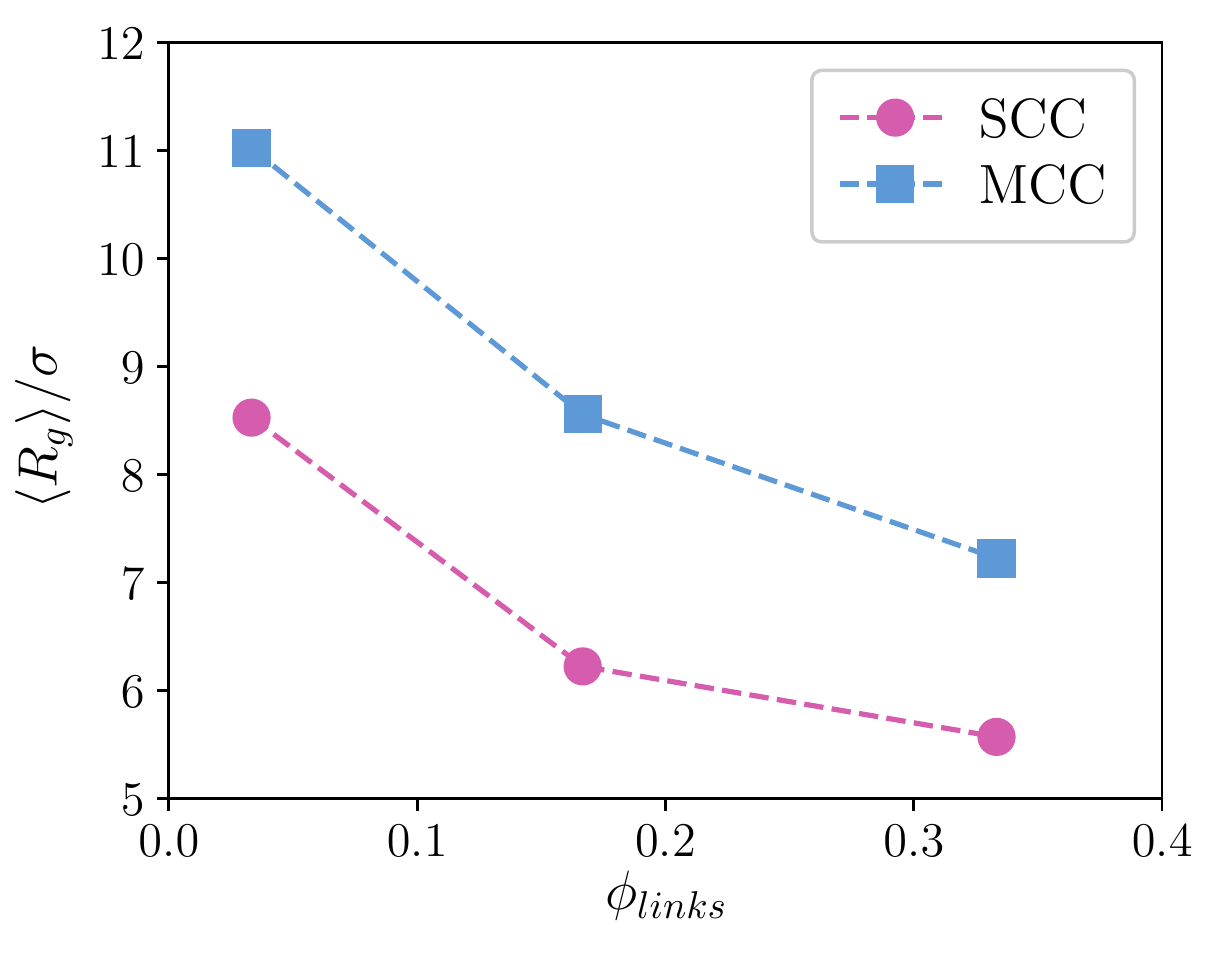}}
 \subfigure[]{\label{fig:Good_size_asph}\includegraphics[width=0.45\textwidth]{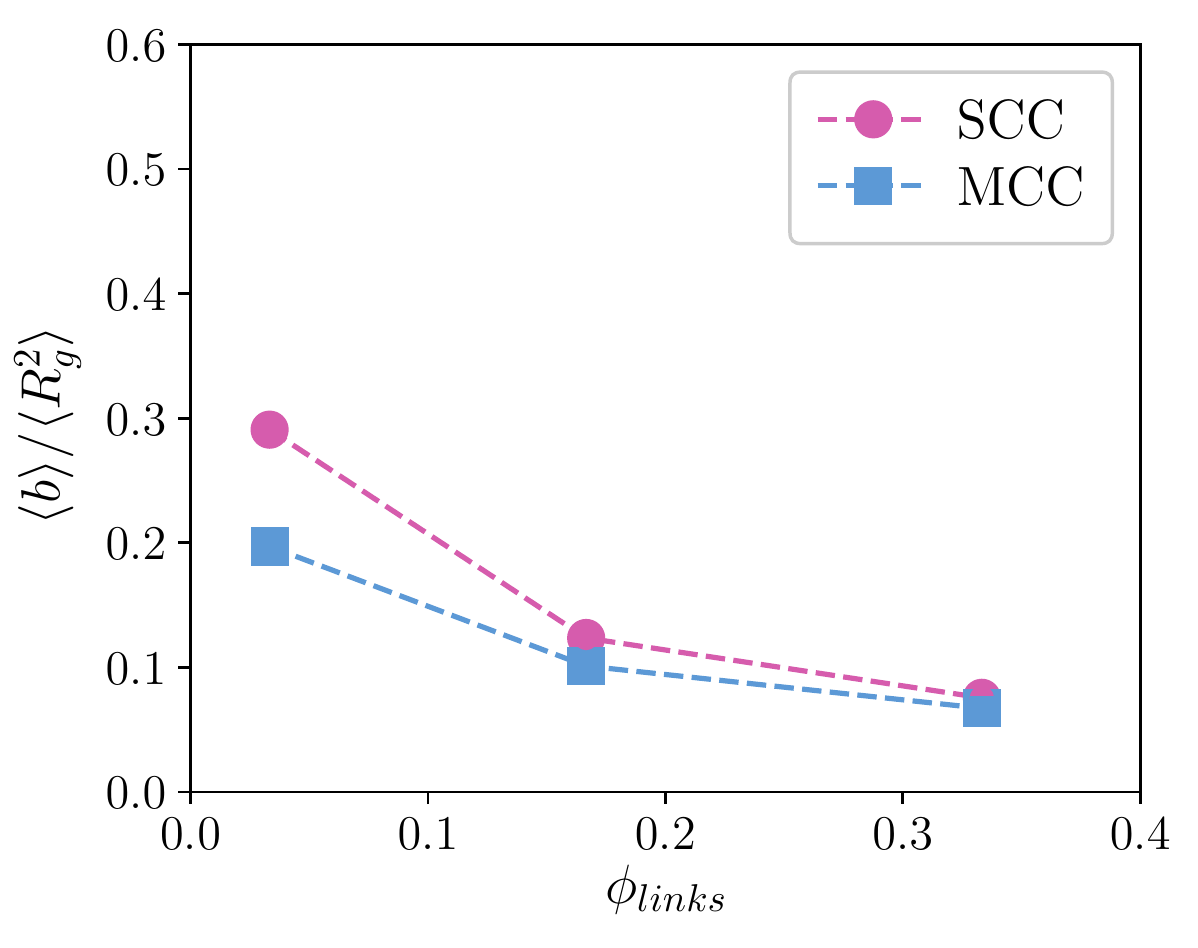}}
 \caption{Reduced average radius of gyration, $\langle R_g \rangle /\sigma$, (a) and average asphericity normalized by the average square radius of gyration, $\langle b \rangle / \langle R_g^2 \rangle $, (b) for every sampled fraction of crosslinks and crosslinking confinement conditions, obtained after a final unconfined equilibration.}\label{fig:Good_size}
\end{figure}

The radius of gyration, $R_g$, and asphericity, $b$, are standard scalar parameters used to characterize the overall structure of soft colloidal particles. The radius of gyration provides the characteristic size of the particle, whereas the asphericity is a non-negative parameter that tends to zero as the shape of the particle approaches a perfect sphere. Both quantities can be computed from the eigenvalues of the gyration tensor, $\{\lambda_1, \lambda_2, \lambda_3\}$ (with $\lambda_1>\lambda_2>\lambda_3$), in the following way:
\begin{equation}
R_g=\left ( \lambda_1^2+\lambda_2^2+\lambda_3^2 \right )^{1/2},
\end{equation}
\begin{equation}
b= \lambda_1^2-\frac{1}{2}\left(\lambda_2^2+\lambda_3^2\right).
\end{equation}
Fig.~\ref{fig:Good_size} shows the average values of the reduced radius of gyration, $\langle R_g \rangle/\sigma$, and the average asphericity normalized by the square radius of gyration, $\langle b \rangle / \langle R_g^2\rangle $, obtained for both system types, as a function of the fraction of crosslinks. In Fig.~\ref{fig:Good_size_radius}, we can observe that, as expected, the size of the particle decreases in all cases when the fraction of crosslinks increases: $\langle R_g \rangle$ decays by approximately $35\%$ when the value of $\phi_{links}$ is increased by 10 times, making the polymer network more compact. Such compaction is also clearly accompanied by the adoption of a more spherical overall profile (see Fig.~\ref{fig:Good_size_asph}). This general behavior is qualitatively consistent with theoretical predictions and experimental measurements of nano- and microgel particles. Importantly, while the case of lowest fraction of crosslinks remains arguable, the compactness and low asphericity of the structures corresponding to the intermediate and the largest fractions of crosslinks supports that, despite their small size, these crosslinked networks can be considered actual nanogel particles.

\subsection{Density profiles}
\begin{figure}[h]
 \centering
 \subfigure[]{\label{fig:Good_DP_100}\includegraphics[width=0.45\textwidth]{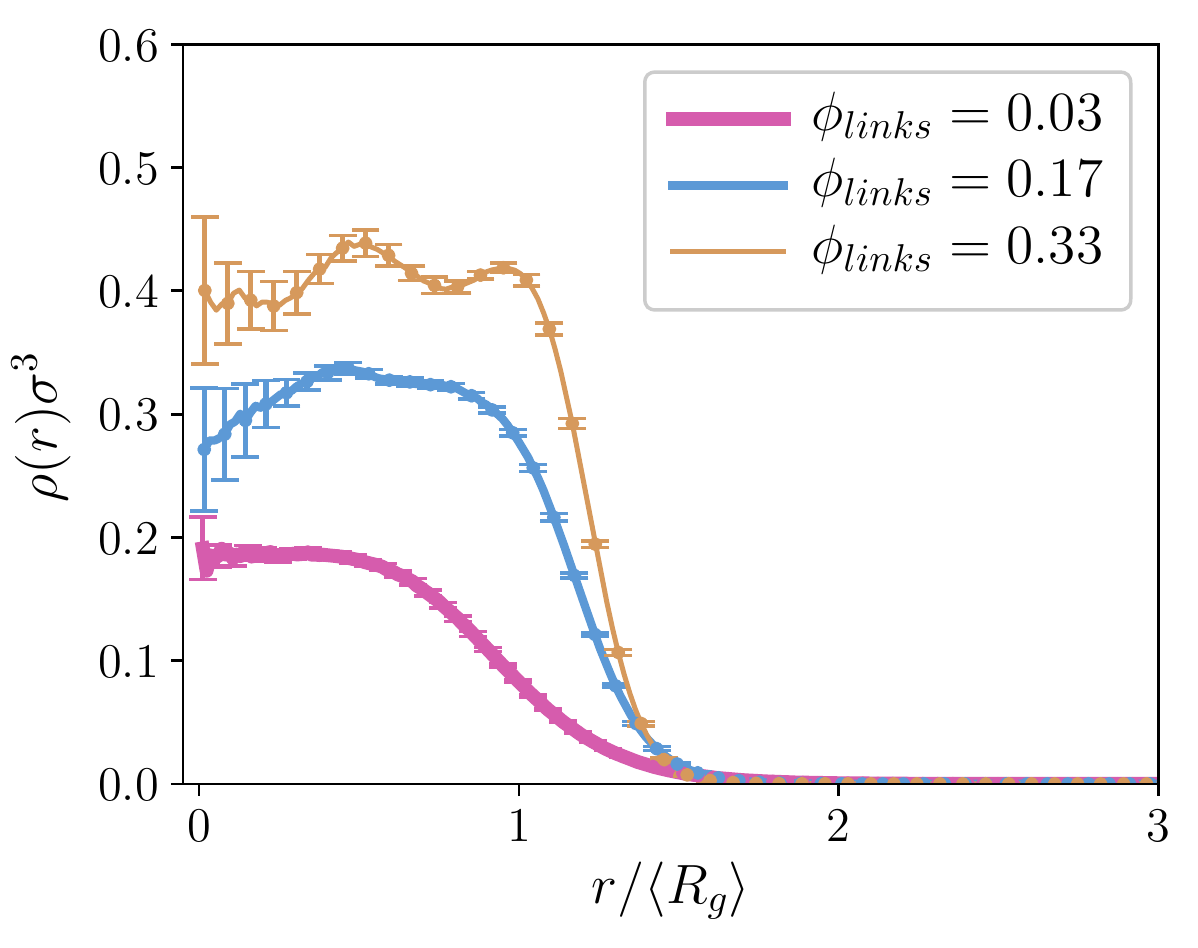}}
 \subfigure[]{\label{fig:Good_DP_200}\includegraphics[width=0.45\textwidth]{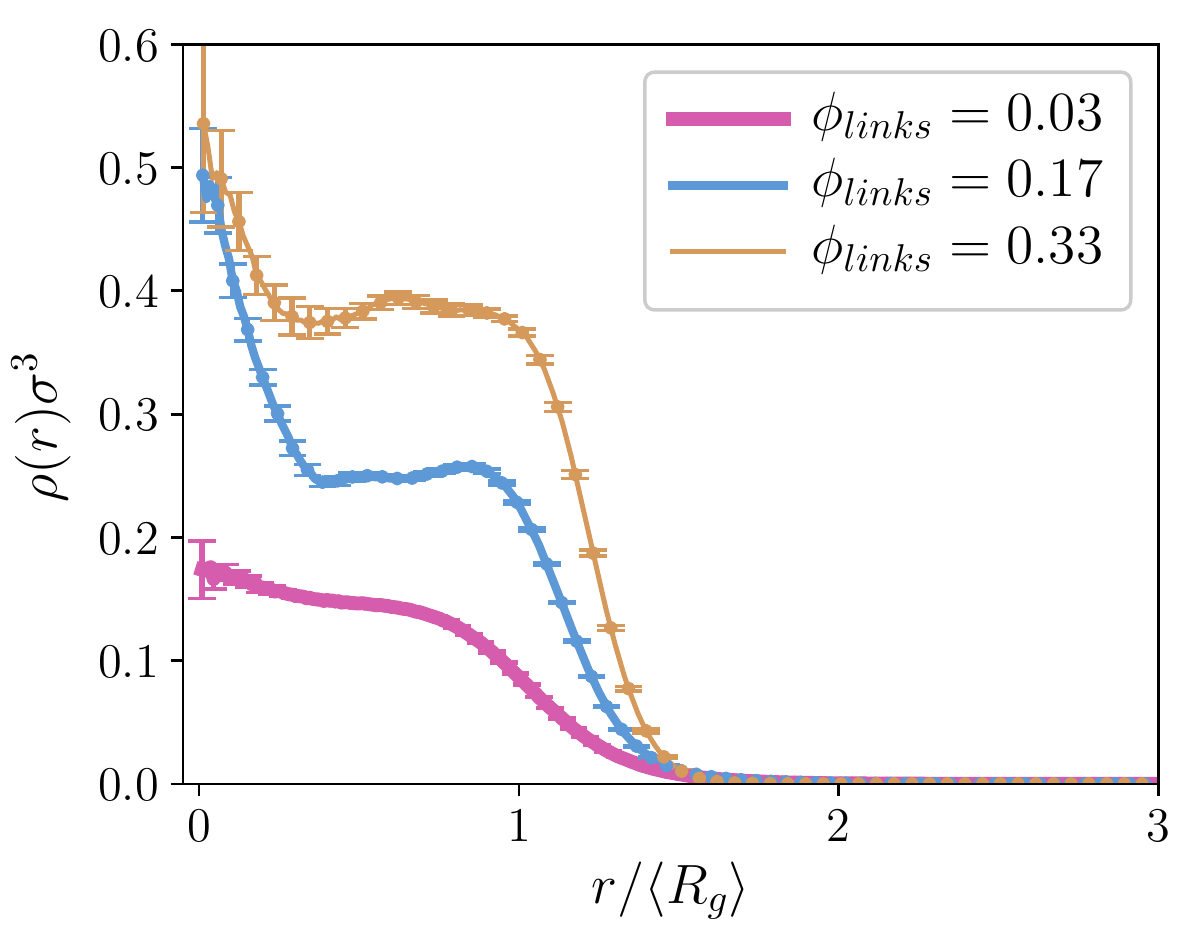}}
 \caption{Reduced number density profiles of polymer beads in SCC (a) and MCC (b) systems, as a function of the normalized distance to the center of mass and corresponding to different fractions of crosslinks, obtained after unconfined equilibration.}\label{fig:Good_DP}
\end{figure}

At this point, we are interested in checking whether the characteristic density profiles observed under different confinement conditions (Fig.~\ref{fig:P_cross-link_100} and \ref{fig:P_cross-link_200}) remain once the networks and nanogel particles are equilibrated without confinement. For this, we compute the corresponding reduced number density profiles, $\rho(r) \sigma^3$, measured in this case as a function of the distance to the center of mass of the system. Fig.~\ref{fig:Good_DP} shows the results obtained for this parameter, plotted for distances normalized with the values of radius of gyration discussed above.
We can observe that, for both system types, profiles of nanogel particles ($\phi_{links}=0.17$ or $0.33$) exhibit a plateau-like region for distances $0.5 \lesssim r/\langle R_g \rangle \lesssim 1.0$ and a relatively abrupt drop at larger distances. For $\phi_{links}=0.03$, however, the drop starts at smaller distances and is significantly smoother. This is due to the relative lack of average internal structure of the latter, especially when compared to nanogel particles. Importantly, the dependence of the central region of the profile on the crosslinking confinement not only persists here for the whole final unconfined system but, in some cases, it is enhanced: for MCC nanogels there is a clear maximum of density at their centers of mass, whereas for SCC ones such central region tends to have a density similar (for high fraction of crosslinks) or lower (for intermediate fraction of crosslinks) than the main outer region. Even the statistical fluctuations at the central region are relatively large, the significant difference between the SCC and MCC cases is clear. For SCC with $\phi_{links}=0.33$, one can also observe a weak indication of layering in the intermediate region of the particle. All these aspects could be important for the design of nanogel particles aimed at applications that require or could benefit from specific internal structuring.

\subsection{Poor solvent conditions}
Structural collapse of colloidal gels, from soft particles with fuzzy boundaries to compact solid-like particles with sharp edges, may happen as a response to certain external stimulus, as for instance changes in solvent quality \cite{2018-keidel-sa}.
\begin{figure}[h]
\centering\includegraphics[width=0.7\textwidth]{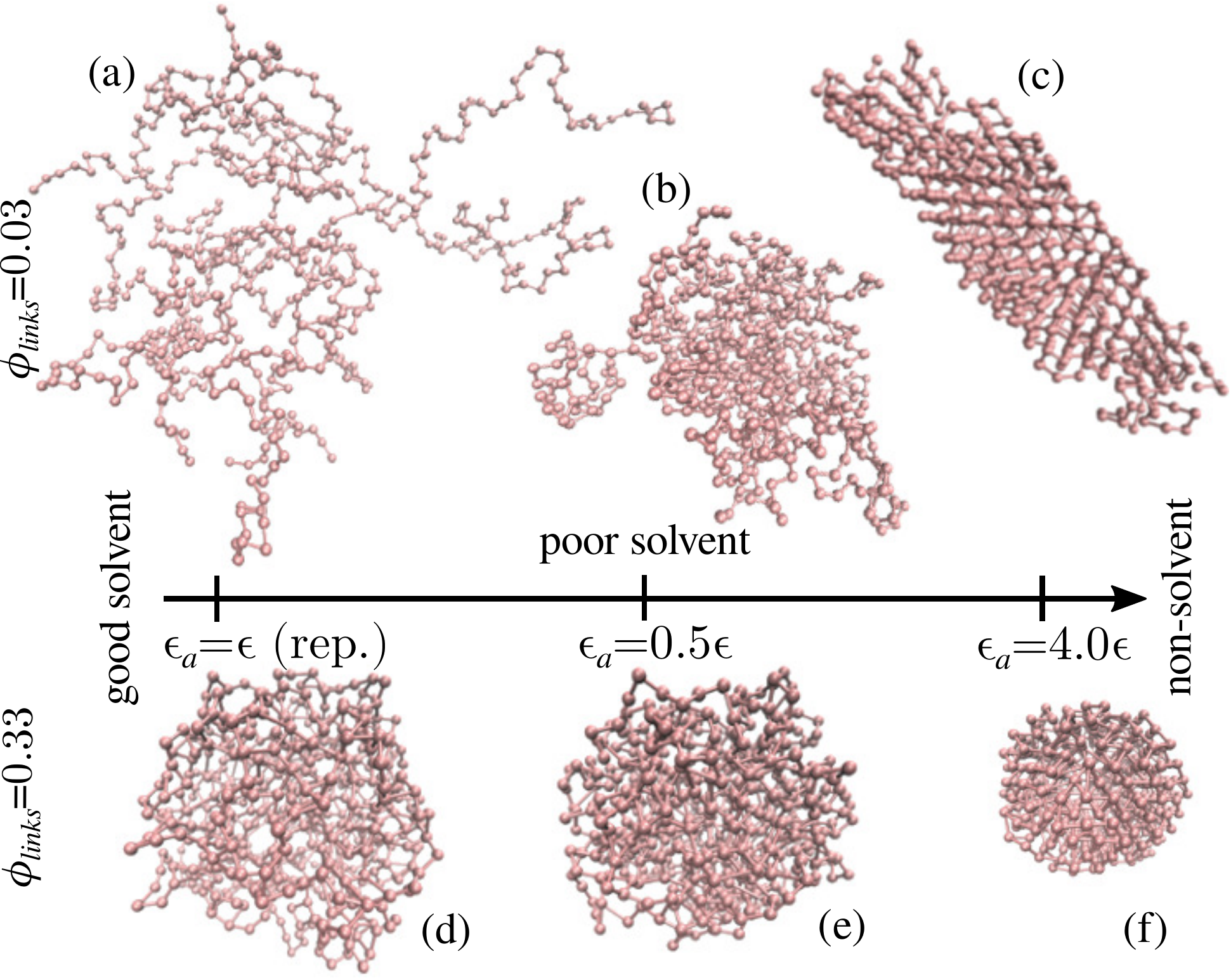}
\caption{Snapshots of unconfined equilibrium configurations of SCC systems with $\phi_{links}=0.03$ (upper row) and $\phi_{links}=0.33$ (lower row), showing the effect of the solvent quality. (a),(d) Good solvent conditions, with only repulsive soft-core interactions between the beads; (b),(e) networks and nanogels shrink in a poor solvent with weak attraction, $\epsilon_{a}=0.5\epsilon$; (c),(f) collapse under non-solvent conditions, corresponding to very strong attraction, $\epsilon_{a}=4.0\epsilon$.}\label{fig:snaphots_attraction}
\end{figure}

Here, we analyze the effects of solvent quality in the structure of our crosslinked networks by making attractive the isotropic soft-core interaction (Eq.~(\ref{eq:WCA})) between polymer beads. Fig.~\ref{fig:snaphots_attraction} shows configuration snapshots that provide two examples of the structural evolution as the strength of the attraction, $\epsilon_a$, is increased. They correspond to SCC systems, with the lowest and highest sampled fractions of crosslinks. In general, the change from a good solvent ($\epsilon_a=\epsilon$ repulsive, left column in Fig.~\ref{fig:snaphots_attraction}) to a poor one ($\epsilon_a=0.5\epsilon$, middle column) leads to the formation of more compact structures, with smaller characteristic size. By further increasing $\epsilon_a$ to reach a very strong attraction---\textit{i.e.}, tending to the limit of no-solvent conditions \cite{Rubinstein03}---the beads are forced to minimize the volume of the structures they form, eventually collapsing into a compact arrangement ($\epsilon_a=4.0\epsilon$, right column in Fig.~\ref{fig:snaphots_attraction}). In our simulations, the transition to a collapsed structure has been observed for $\epsilon_a$ in the range from $2.0\epsilon$ to $2.5\epsilon$. Fig.~\ref{fig:snaphots_attraction} also shows the existence of a particular case in the way the networks tend to collapse: whereas in all other cases, we observed that the structures tend to become compact spherical particles (lower row in Fig.~\ref{fig:snaphots_attraction}) as reported in numerous studies, in the case of the polymer network formed with the lowest fraction of crosslinks (upper row in Fig.~\ref{fig:snaphots_attraction}) the collapsed structure is very anisometric, approaching a cylindrical shape, as a consequence of the low amount of crosslinks and their highly irregular distribution. This is another indication that such loose networks do not fulfill the common characteristics of nanogel particles.

\begin{figure}[h]
 \centering
 \subfigure[]{\label{fig:RG_attractive_100}\includegraphics[width=0.45\textwidth]{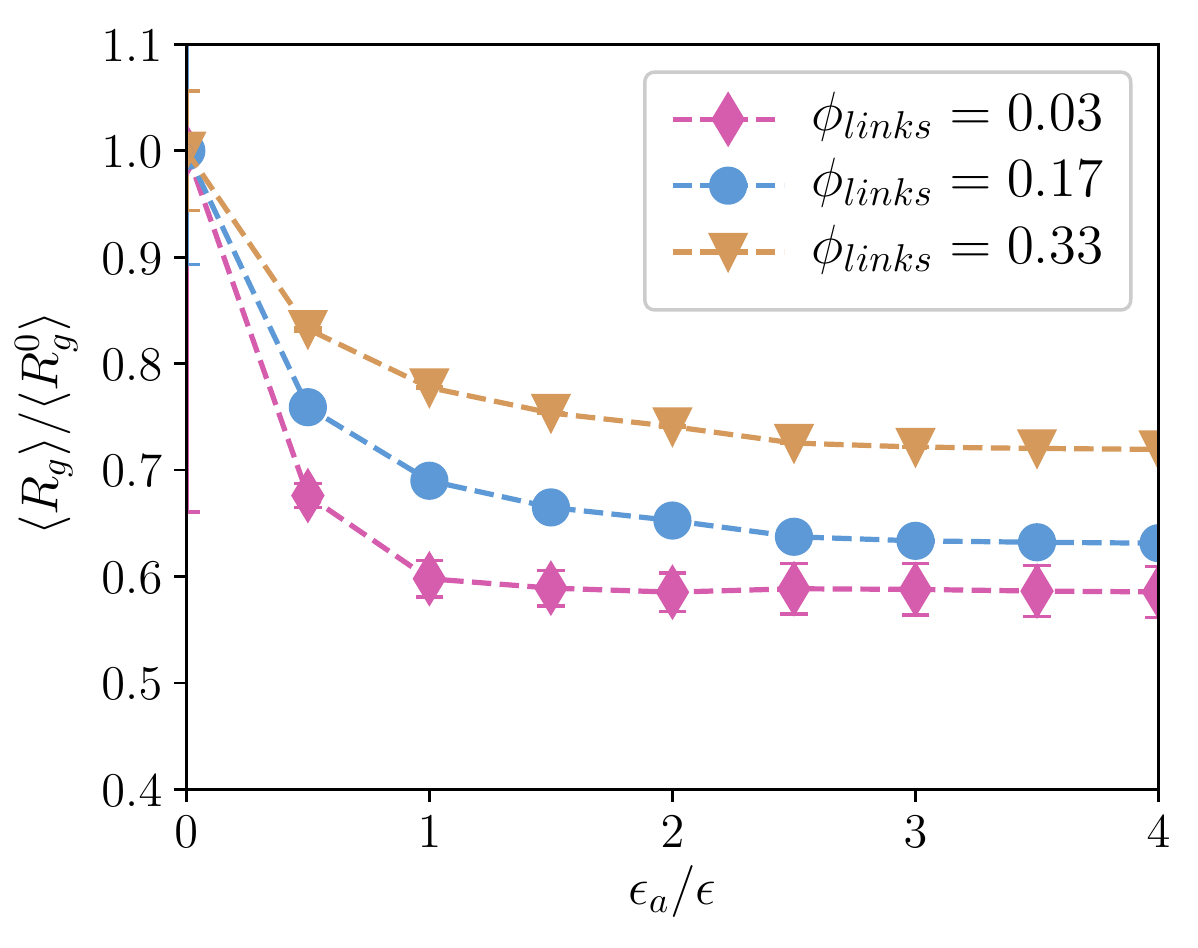}}
 \subfigure[]{\label{fig:RG_attractive_200}\includegraphics[width=0.45\textwidth]{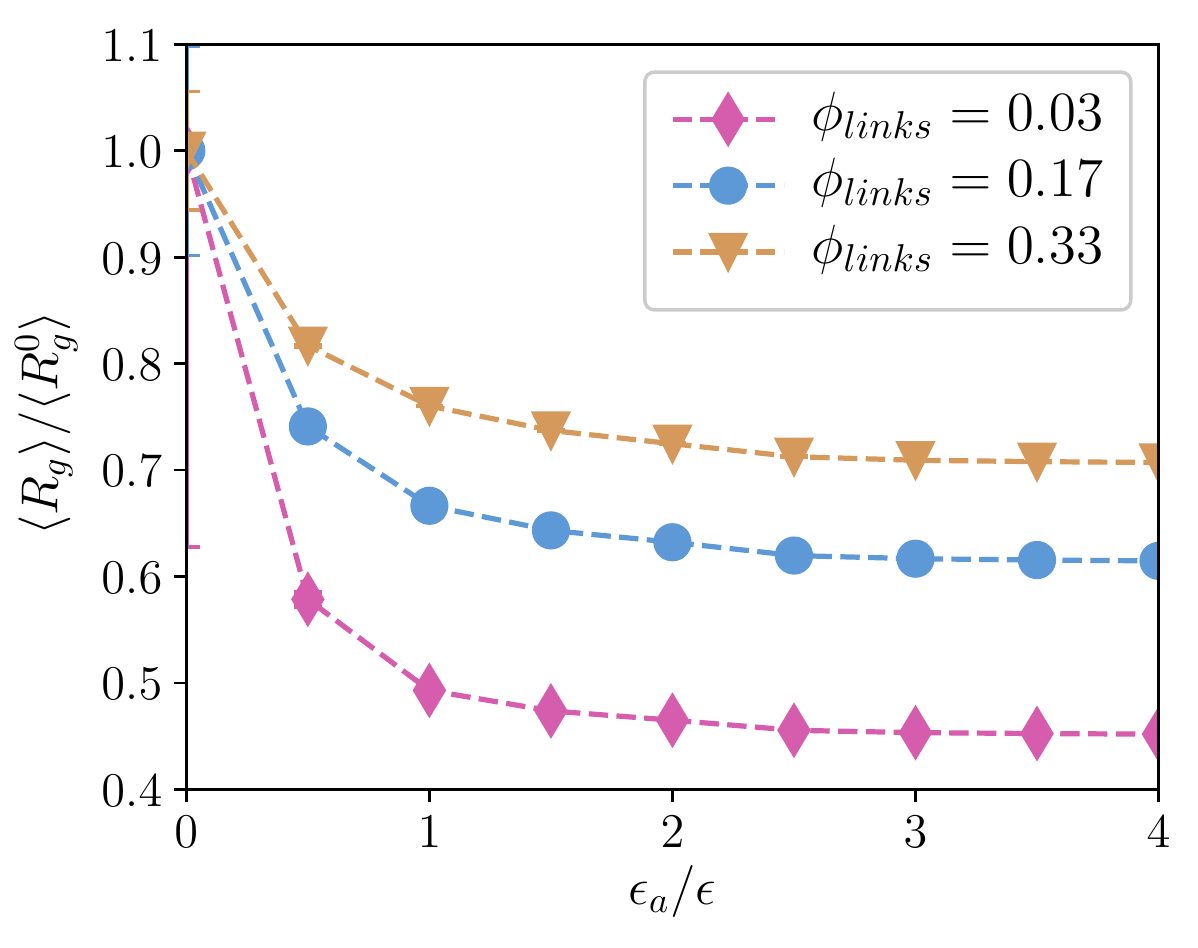}}
 \caption{Changes in the average radius of gyration, $\langle R_g \rangle$, as a function of the reduced attraction strength corresponding to poor solvent conditions, $\epsilon_a/ \epsilon$, and relative to its value under good solvent conditions, $\langle R_g^0\rangle$, for different fractions of crosslinks. (a) SCC systems. (b) MCC systems.}\label{fig:RG_attractive}
\end{figure}
\begin{figure}[h]
 \centering
 \subfigure[]{\label{fig:asphericity_attractive_100}\includegraphics[width=0.45\textwidth]{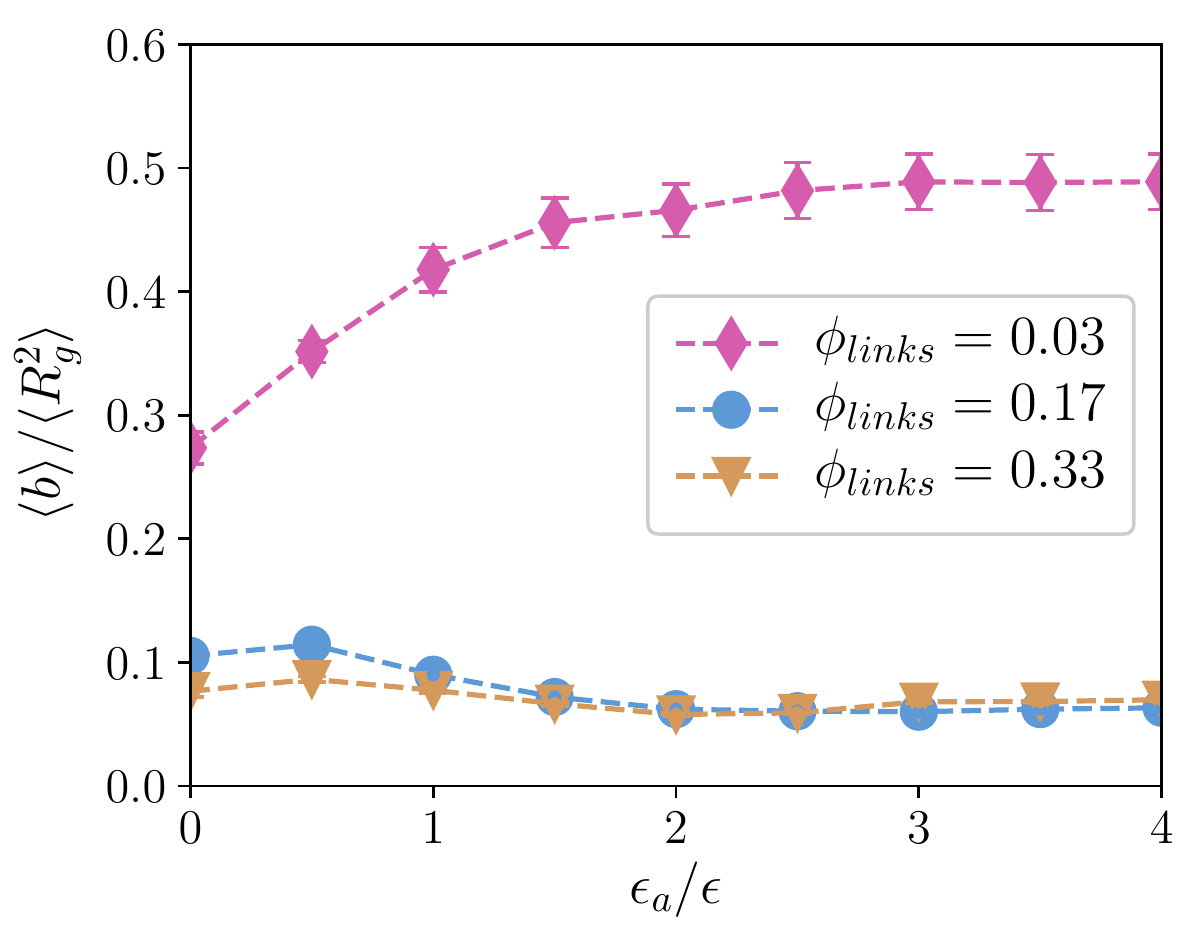}}
 \subfigure[]{\label{fig:asphericity_attractive_200}\includegraphics[width=0.45\textwidth]{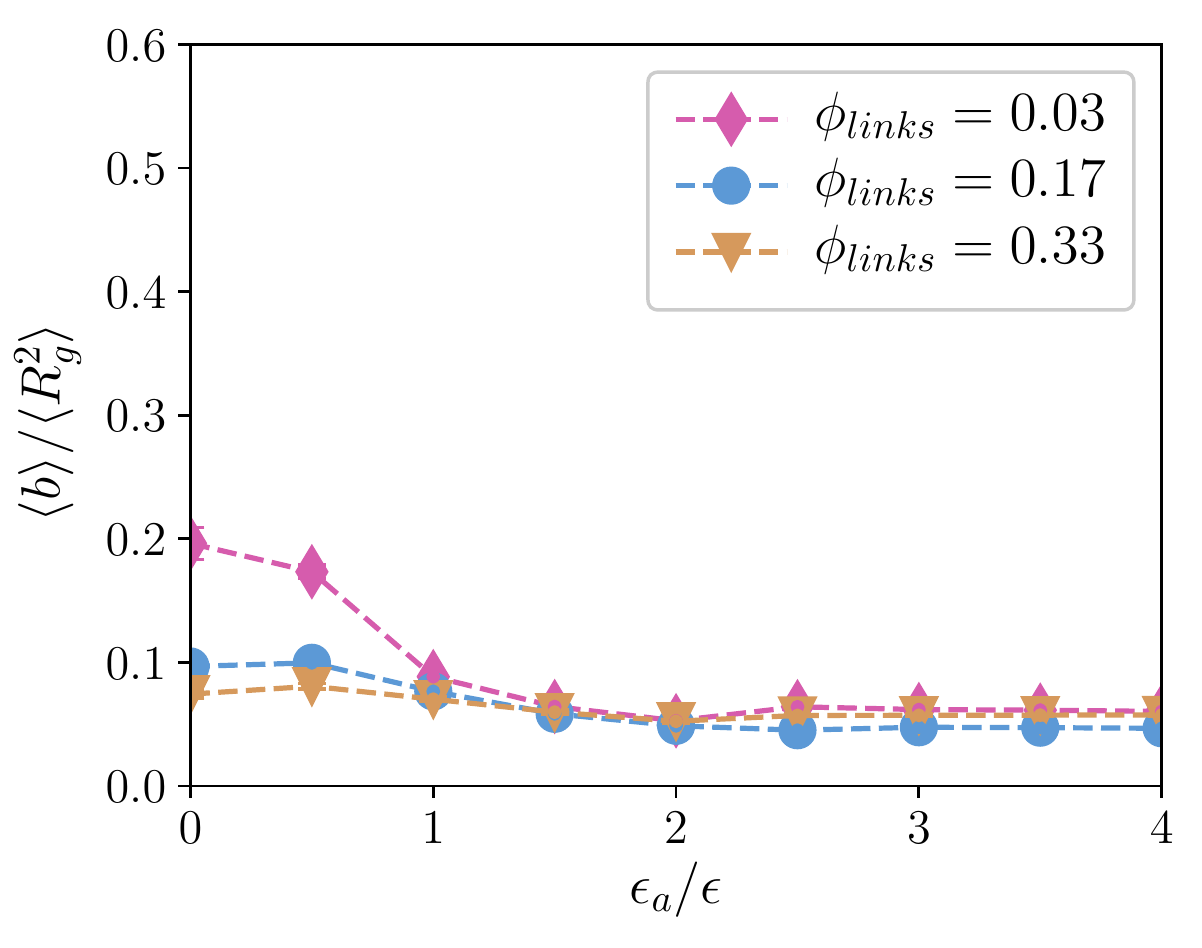}}
 \caption{Average asphericity scaled with the average square radius of gyration, $\langle b \rangle/ \langle R_g^2 \rangle$, as a function of the attraction strength associated to poor solvent conditions, for different fractions of crosslinks. (a) SCC systems. (b) MCC systems.}\label{fig:asphericity_attractive}
\end{figure}

The qualitative structural evolution observed in the snapshots can be better characterized by means of the radius of gyration and the asphericity. The results obtained for these parameters as a function of the attraction strength are presented in Fig.~\ref{fig:RG_attractive} and \ref{fig:asphericity_attractive}, respectively. Specifically, Fig.~\ref{fig:RG_attractive} shows the average reduced radius of gyration, $\langle R_g \rangle/\sigma$, relative to the value corresponding to good solvent conditions, $\langle R_g^0 \rangle =\langle R_g(\epsilon_a=\epsilon,\,\mathrm{repulsive}) \rangle$, for different fractions of crosslinks. From these plots, one can conclude that $\langle R_g \rangle/\sigma$ decays abruptly as soon as solvent quality worsens. This decrease tends to saturate once the attraction strength becomes more important than the thermal fluctuations, $\epsilon_a/\epsilon>1$. In general, this relative decay is larger for MCC system, reflecting its initial lower overall density. In addition, the strongest decay is observed for the most loosely crosslinked system, as a consequence of the strong nonuniform shrinking from a rather expanded network.

The effect of the internal structure also manifests itself in the normalized average asphericity, $\langle b \rangle / \langle R_g^2\rangle$, shown in Fig.~\ref{fig:asphericity_attractive}. In agreement with the discussion above on the snapshots of Fig.~\ref{fig:snaphots_attraction}, in all cases except for the most loosely crosslinked SCC system, the asphericity drops around $\epsilon_a/\epsilon \approx 1$ to become very small for $\epsilon_a/\epsilon>1$, indicating the rather spherical shape of the particles under such conditions. The exceptional case corresponding to a strongly anisometric collapsed configuration is reflected in a large growth of asphericity with $\epsilon_a/\epsilon$ (Fig.~[\ref{fig:asphericity_attractive_100}]).

\subsection{Ionic systems}
An important part of nano- and microgels created to date consist of charged polymers that introduce large structural responses to variations in the ionic strength of the solvent, particularly strong swelling and deswelling behaviors. These ionic systems have been the subject of numerous theoretical studies, especially by means of computer simulations based on regular lattices~\cite{Claudio2009,Kobayashi2014,Ghavami2016,Kobayashi2016,Ahualli2017,Ghavami2017,Kobayashi2017,Hofzumahaus2018}. Here, we also perform a preliminary study of the properties of our model nanogel particles and networks when synthesized from polyelectrolyte precursors, using the Debye-H\"uckel approximation (Eq. (\ref{eq:DH})). We stress that, here, we are interested only in a qualitative analysis of the structural changes led by electrostatic interactions and their dependence on the different network topologies produced by distinct confinement conditions during synthesis. Therefore, we sample reduced parameters aimed at underlining such changes rather than at the accurate representation of the system. Specifically, here we assign to each polymer bead a charge of $q_i=-1$ and sample values of the electrostatic interaction strength, $\epsilon_{\kappa}/\sigma \epsilon$, in the range $\epsilon_{\kappa}/\sigma \epsilon \in [0.5, 4.0]$ for two values of the inverse Debye screening length, $\kappa\sigma=0.4$ and $0.7$.

\begin{figure}[!t]
\centering
\includegraphics[width=0.8\textwidth]{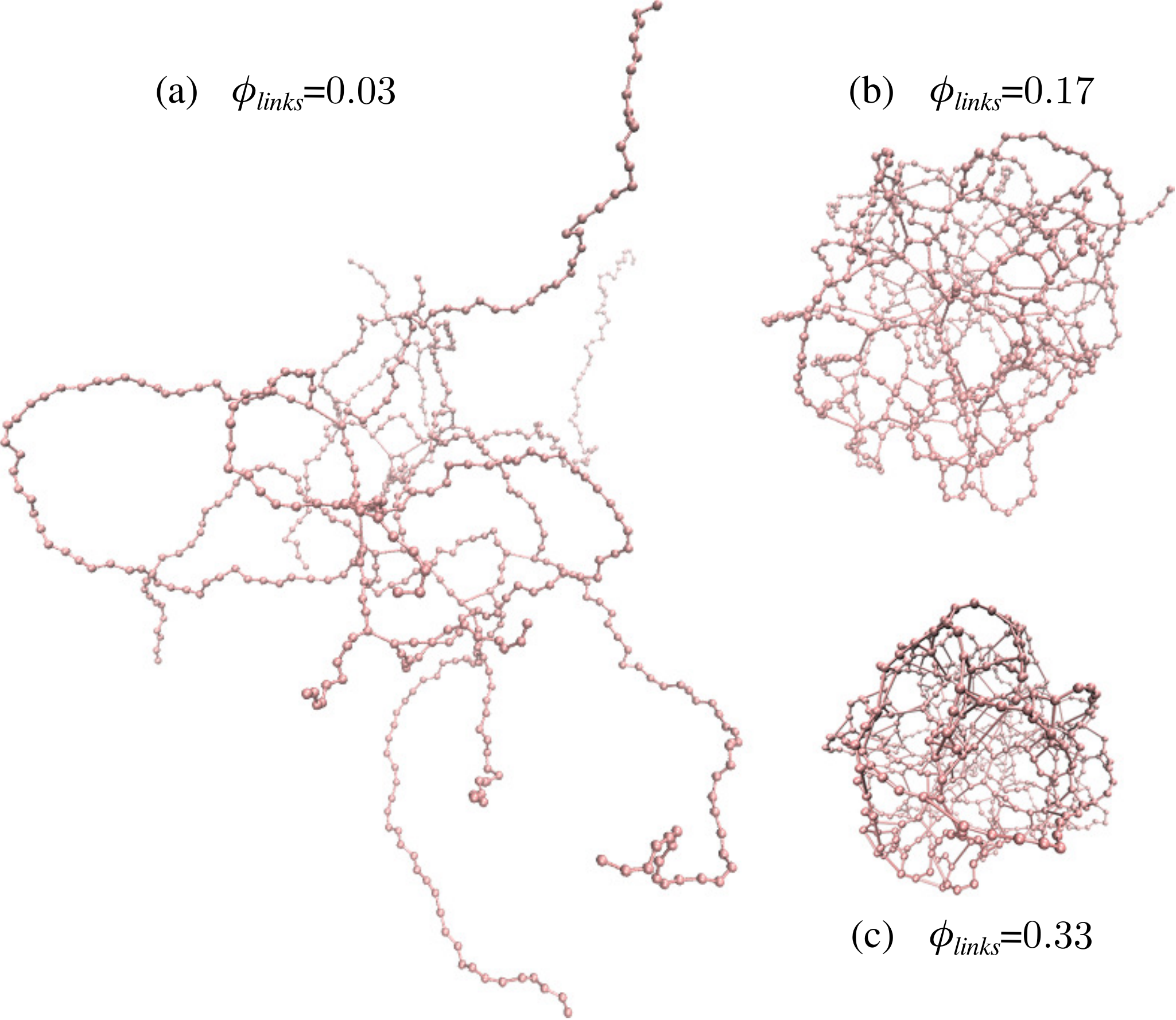} 
\caption{Simulation snapshots of SCC systems with swollen structures due to Debye-H\"uckel electrostatic interactions (with $\epsilon_{\kappa}/\sigma\epsilon=4.0$, $\kappa\sigma=0.4$) and different fractions of crosslinks.}\label{fig:snapshots_yukawa}
\end{figure}
Fig.~\ref{fig:snapshots_yukawa} includes configuration snapshots of systems analogous to the ones presented in Fig.~\ref{fig:snapshots_good}, that show a clear swelling due to a strong, slightly screened electrostatic repulsion between the beads. This leads to a stiffening of the linear segments in the polymer networks, causing the overall expansion of the structure. By comparing both sets of snapshots, we can see that loosely crosslinked networks ($\phi_{links}=0.03$) show a very strong expansion, loosing any eventual particle-like shape. However, despite still experiencing a considerable swelling, nanogel particles ($\phi_{links}=0.17,\,0.33$) keep their nearly spherical shape.
\begin{figure}[t]
 \centering
 \subfigure[]{\label{fig:Rg_yukawa_100}\includegraphics[width=0.45\textwidth]{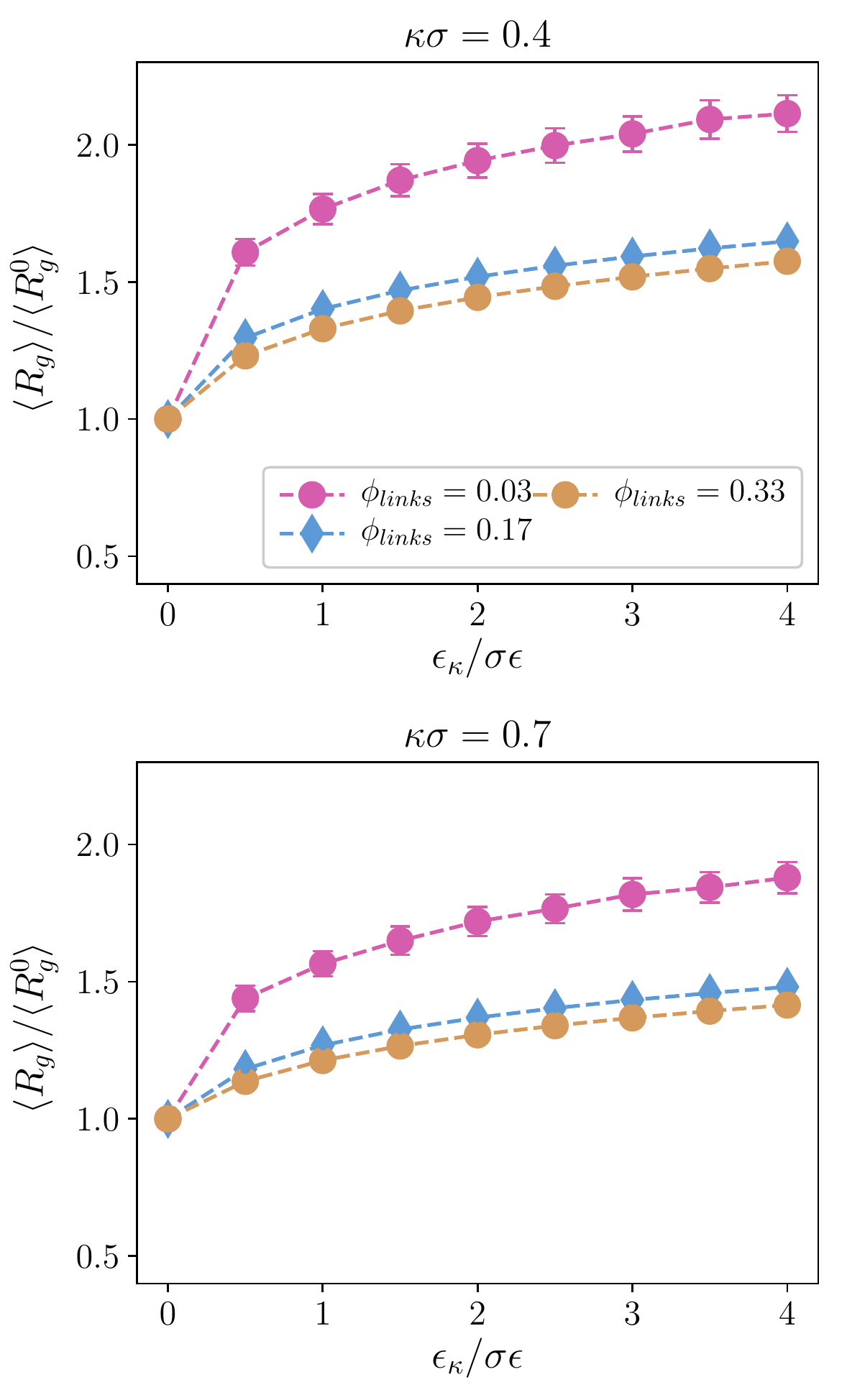}}
 \subfigure[]{\label{fig:Rg_yukawa_200}\includegraphics[width=0.45\textwidth]{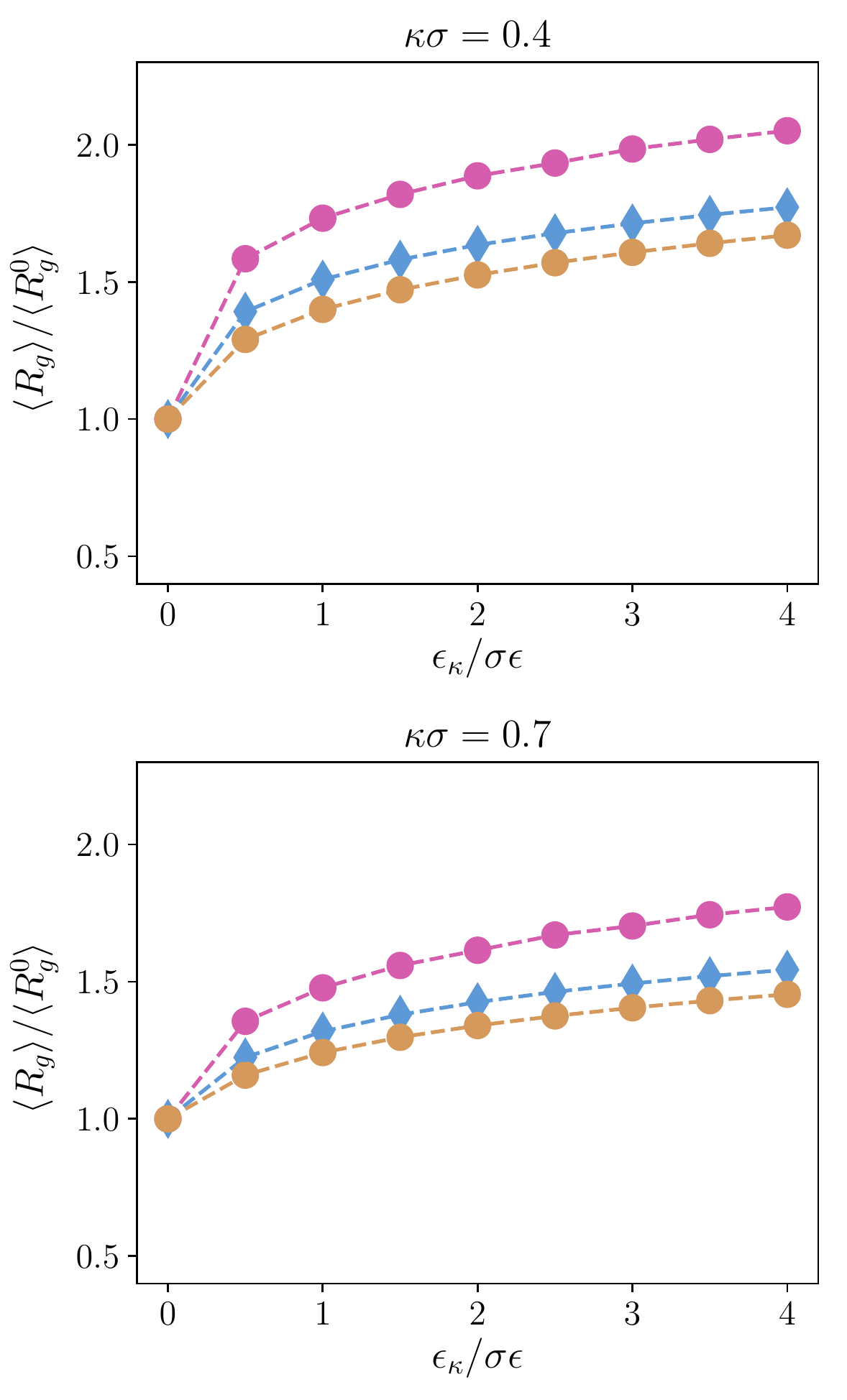}}
 \caption{Relative change in the average radius of gyration, $\langle R_g\rangle$, as a function of the reduced strength of the electrostatic interactions, $\epsilon_{\kappa}/\sigma\epsilon$, with respect to the non-ionic case, $\langle R_g^0 \rangle$, for different fractions of crosslinks. Upper row corresponds to a Debye screening wavelength of $\kappa\sigma = 0.4$, lower row to $\kappa\sigma = 0.7$. Left column: SCC systems. Right column: MCC systems.}\label{fig:Rg_yukawa}
\end{figure}
Fig.~\ref{fig:Rg_yukawa} illustrates in more detail, also by means of the relative average radius of gyration, the swelling experienced by systems of each type and fraction of crosslinks as the strength of the electrostatic interactions increases. As expected, the largest relative expansion (above $100\%$ within the sampled range) happens for the system that was proven to be structurally the weakest---\textit{i.e.}, the SCC with lowest $\phi_{links}$---under the lower screening conditions, $\kappa \sigma =0.4$ (see Fig.~\ref{fig:Rg_yukawa_100}, top row). The smallest expansion corresponds to MCC systems with largest $\phi_{links}$ and screening (around $45\%$, see lower row in Fig.~\ref{fig:Rg_yukawa_200}). Also important is that, confirming the observation above, the average asphericity of the nanogel particles (not shown) does not change significantly with their electrostatically-induced swelling.
\begin{figure}[h]
 \centering
 \subfigure[]{\includegraphics[width=0.45\textwidth]{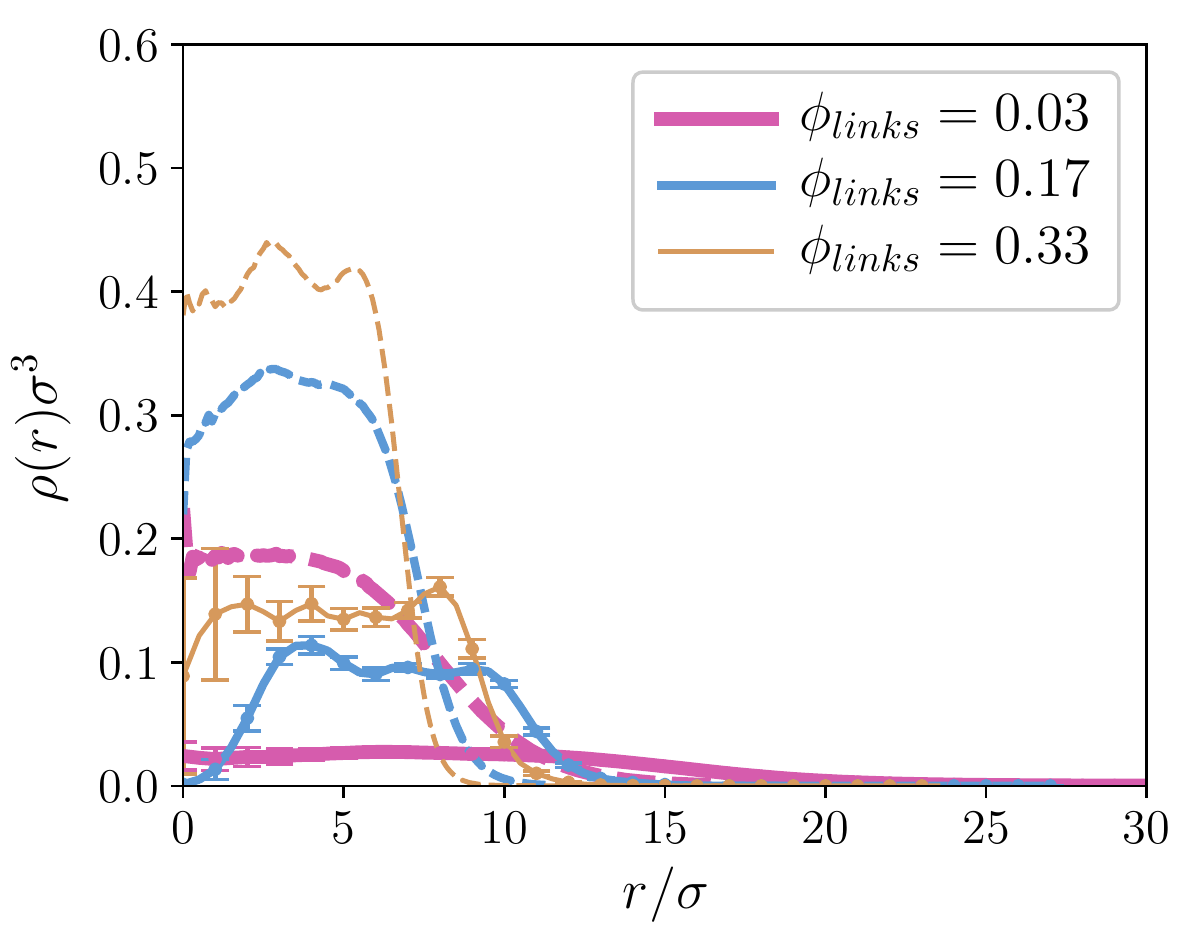}}
 \subfigure[]{\includegraphics[width=0.45\textwidth]{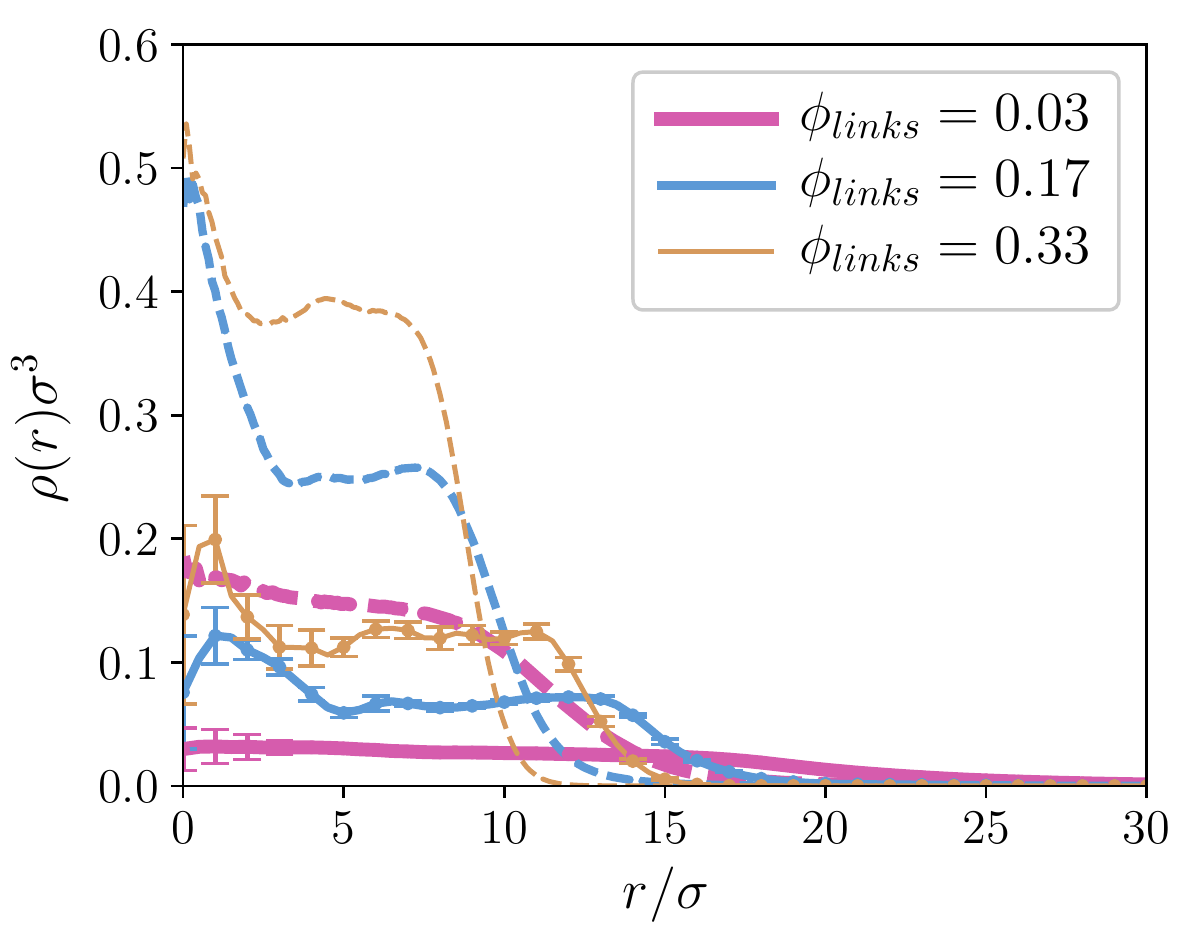}}
 \caption{Reduced number density profiles at different fractions of crosslinks, corresponding to non-ionic (dashed lines) and ionic systems (solid lines) with $\epsilon_{\kappa}/\sigma\epsilon=4.0$ and $\kappa\sigma=0.7$. (a) SCC systems. (b) MCC systems.}\label{fig:Yukawa_DP}
\end{figure}

Finally, we check the changes in the internal structure due to the electrostatic swelling by comparing the density profiles of the non-ionic systems, already introduced in Fig.~\ref{fig:Good_DP}, with the ionic ones. An example of this comparison, corresponding to the case of strong electrostatic interactions and screening, is shown in Fig.~\ref{fig:Yukawa_DP}. We can see that, besides the significant broadening of the profiles, the main qualitative features observed for non-ionic systems remain in their ionic counterparts: a decay in the external region of the profiles, more or less smooth depending on the fraction of crosslinks; a plateau-like intermediate region and a central region where the differences in the internal topology led by the original confinement conditions manifest. Interestingly, the drop in density in the central region that corresponds to SCC nanogels with intermediate fraction of crosslinks is significantly enhanced by the electrostatic swelling, showing that the internal structure is basically preserved despite the general expansion of the interstitial regions. In turn, this type of expansion also makes the maximum in density near the center of mass of the larger nanogel particles to become much less prominent. Therefore, we can conclude that electrostatic interactions basically tend to enhance the particular features of the internal structure of nanogels obtained under strongly confined moderate crosslinking.

\subsection{Form factor measurements}
As a last result, we check whether the differences in the internal structure we found in our crosslinked networks and nanogel particles depending on the confinement conditions might be detected in scattering experiments. Note that here, we do not aim at a formal characterization of small angle scattering data profiles, but only at providing a qualitative hint on the experimental feasibility of such analysis.

The component that contains the information corresponding to single particles in scattering measurements data is known as form factor. SANS measurements are frequently employed to deduce the form factors of micro- and nanogels particles, usually by means of fitting the experimental data to a given predefined model of the internal structure.\cite{2014-schneider-lm, 2016-felberg-mph, 2019-witte-sm} In simulation data, one can calculate form factors directly from the coordinates of the elements of the particle. In our case, the following expression can be applied to each pair of polymer beads in the network, $i$ and $j$:
\begin{figure}[t]
 \centering
 \subfigure[]{\includegraphics[width=0.45\textwidth]{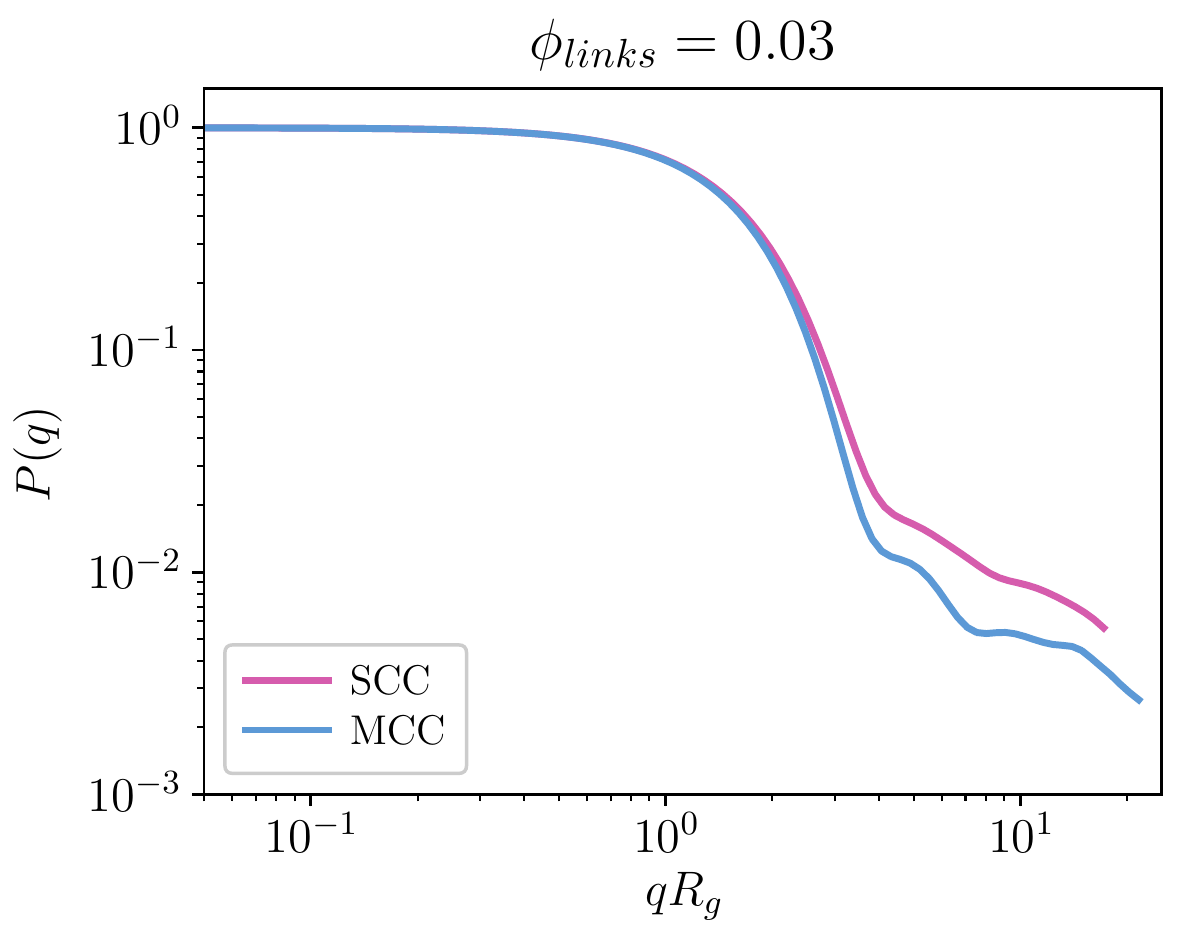}}
 \subfigure[]{\includegraphics[width=0.45\textwidth]{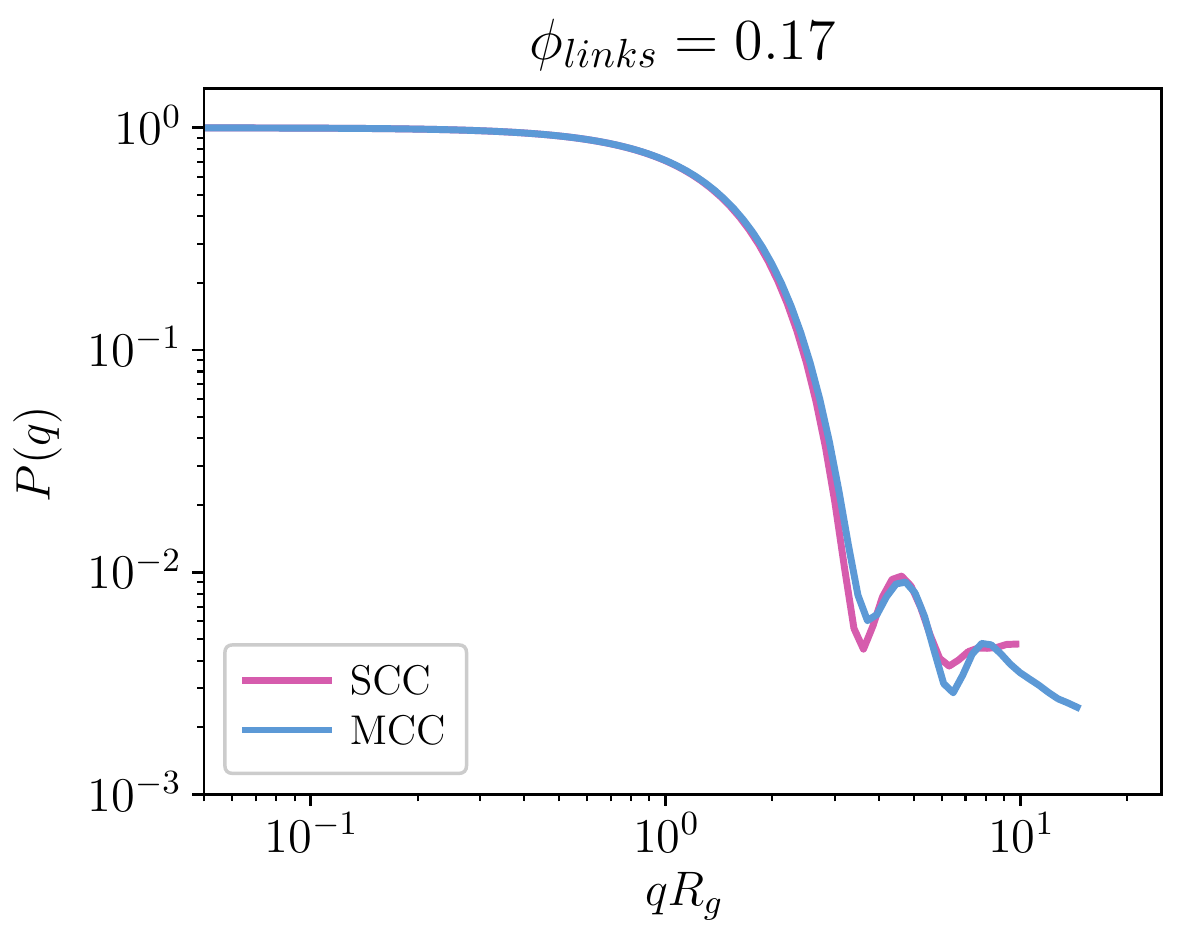}}
 \subfigure[]{\includegraphics[width=0.45\textwidth]{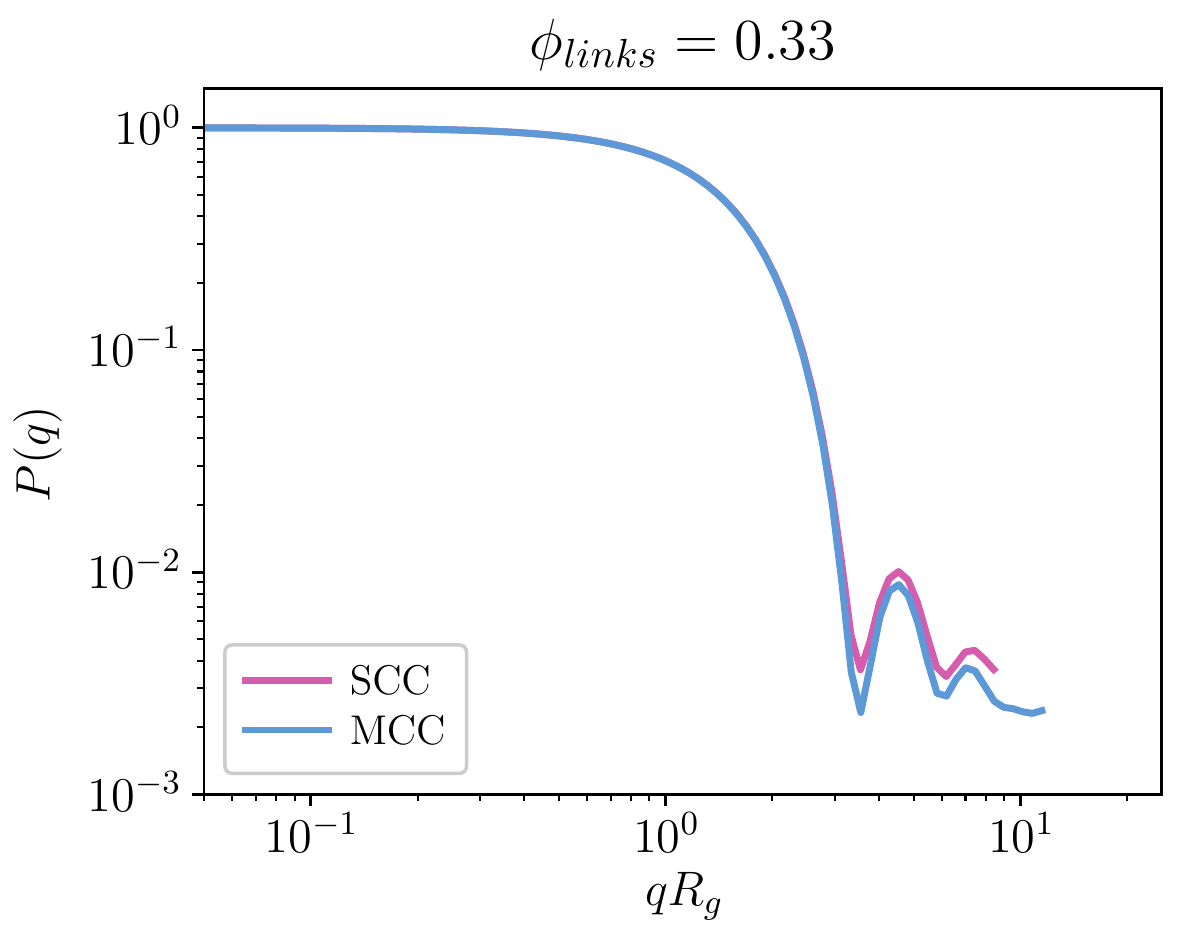}}
 \caption{Comparison of form factors, $P(q)$, of ionic systems with $\epsilon_{\kappa}/\sigma \epsilon=4.0$, $\kappa\sigma=0.4$ and different fractions of crosslinks, for systems corresponding to strong (SCC) and moderate (MCC) confinement conditions during crosslinking.}\label{fig:form_factors}
\end{figure}
\begin{equation}
    P(q)=\frac{1}{N^2}\sum\limits_{i=1}^N\sum\limits_{j=1}^N\left\langle\frac{\sin(qr_{ij})}{qr_{ij}}\right\rangle,
    \label{eq:form_factor}
\end{equation}
where $r_{ij}$ is the center-to-center distance between them, $q=\| \vec q \|$ is the length of the scattering wave vector and the angular brackets stand for an averaging over different configurations. Fig.~\ref{fig:form_factors} shows a selection of form factors corresponding to ionic systems under electrostatic swelling conditions ($\epsilon_{\kappa}/\sigma \epsilon=4.0$ and $\kappa\sigma=0.4$), as these are the ones that exhibit the largest quantitative differences in their internal structure. In these plots, the length of the wave vector is rescaled by the corresponding radius of gyration to check that the Guinier regime is established at $q<R_g$.\cite{Guinier1955} This corresponds to the plateau observed for $q R_g < 1$. The difference in the structure of the polymer networks has to manifest in the large $q$ region. For such region we set the limit $q\leq 1/\sigma$, so that we do not reach shorter length scales than that of a single polymer bead diameter, $\sigma$. By comparing the form factors corresponding to moderately and strongly confined crosslinking conditions, in the region of large $q$ one can observe quantitative differences in their profiles, particularly for systems of loosely crosslinked networks. The latter case shows a rather smooth profile that is frequently observed in experimental measurements as a consequence of the smearing produced by a large configurational entropy in the system. In our case, this smearing comes from the averaging over independent realizations. The profiles corresponding to the intermediate and the largest fraction of crosslinks, however, show few clear maxima. This indicates that the internal structure of these nanogel particles, even being formed in a random crosslinking process, keeps a rather consistent distribution among independent realizations. This relatively low internal configurational variation is also favored by the small size of the particles. Therefore, these results suggest that the effects of the confinement conditions during random crosslinking of loose polymer networks and nanogel particles actually might be detected in small angle scattering experimental measurements.

\section{Conclusions}
By means of a coarse-grained, bead-spring computer model of polymers, we studied the effects of strong and moderate confinement conditions on the internal structure and properties of nanogel particles synthesized by random crosslinking of polymer molecules diluted in nanodroplet emulsions. Random crosslinking, that corresponds to current experimental irradiation- or electrochemically-induced crosslinking methods, makes the resulting polymer networks to not have an architecture directly predefined by their chemical initial components. Instead, our results indicate that such internal structure can be influenced by the relative size of the confining droplet.

Our simple random crosslinking model is able to produce polymer networks that, except for a very low fraction of crosslinks, show the characteristic behavior of actual nanogel particles, including structurally stable nearly spherical shapes. In order to understand the role of confinement before and during crosslinking, we compared two scenarios: first, a moderate density of polymer beads inside the confining droplet, or moderate confinement conditions, that makes polymer molecules tend to mix well and have a typical density profile, with a higher density in the center of the droplet, a relatively homogeneous intermediate region and a strong decay at the edge; second, a strong confinement due to a high density of polymer beads, that makes polymers tend to segregate and occupy preferably the region near the confinement boundary, leaving a region of slightly lower density in the center. We observed that random moderate crosslinking preserves such internal structural features in the resulting nanogel particles, which possess different network topologies as revealed by their distributions of linear segments and average shortest paths between crosslinked spots. In particular, a more inhomogeneous distribution of crosslinks is obtained for the crosslinking under stronger confinement conditions.

A qualitative similar compaction and final collapse into compact spherical structures as the quality of the solvent worsens has been also observed in both types of model nanogel particles. Only a collapse into very anisometric structures has been found in the extreme case of polymer networks obtained by the combination of a low fraction of crosslinks and a relatively high inhomogeneity in their distribution due to strongly confined crosslinking conditions.

The usual changes in overall size associated to the swelling/deswelling behavior due to electrostatic interactions have been observed when we consider the nanogel particles to consist of charged polymers. However, the impact on the internal structure is different depending on the topology of the polymer networks. Whereas the density profiles of nanogels synthesized under moderate confinement conditions tend to flatten with the electrostatic swelling, in the ones corresponding to nanogels obtained under strong confinement and moderate degree of crosslinking the central region of lower density tends to be enhanced.

In all aspects analyzed here, the most interesting features have been observed for a moderate value of fraction of crosslinks. This is due to the fact that such parameter controls the tradeoff between the structural stability, that makes the network to have a persistent shape and behave as a soft spherical particle, and the existence of internal structural inhomogeneities, that can be useful or even essential for given practical applications.

In summary, these results suggest the control of confinement conditions, for instance by means of membrane emulsification techniques, as a simple approach to tune qualitatively the structure of nanogel particles using random crosslinking methods, whereas the fraction of crosslinks can be used to tune quantitatively their properties.

As a final outlook comment, in our ongoing research, we extend the current model to consider other sources of internal heterogeneities, able to produce well differentiated core-shell nanogel structures.

\section*{Acknowledgements}

E.~S.~M. and S.~S.~K. acknowledge the financial support of the Austrian Science Fund (FWF): START-Projekt Y627-N27. S.~S.~K. and C.~N.~L. acknowledge the support of ETN-COLLDENSE (H2020-MSCA-ITN-2014, Grant No. 642774). P.~A.~S. is supported by the Act 211 of the Government of the Russian Federation, contract No.~02.A03.21.0006. Computer simulations were performed at the Vienna Scientific Cluster (VSC). We thank Eugenia Kumacheva for fruitful discussions concerning the experimental relevance of our model.

\end{document}